\documentclass[12pt,a4paper]{article}%
\pdfoutput=1
\usepackage{amsmath,amssymb,amsfonts}
\usepackage{caption2}
\usepackage[]{hyperref}
\usepackage[pdftex]{color,graphicx}
\usepackage{multicol}

\numberwithin{equation}{section} 
\setcaptionmargin{1cm}
\newcommand{\hhref}[1]{\href{http://arxiv.org/abs/#1}{{\it arXiv:#1}}}

\newcommand{\p}{\partial}
\def\beq{\begin{equation}}
\def\eeq{\end{equation}}
\def\eq{\end{equation}}
\newcommand{\be}{\begin{equation}}
\newcommand{\ee}{\end{equation}}
\newcommand{\bea}{\begin{eqnarray}}
\newcommand{\eea}{\end{eqnarray}}

\begin{document}


\begin{titlepage}
$\quad$
\vskip 4.0cm
\begin{center}
{\huge \bf  A naturally light dilaton } 
\\ 
\vskip 2.0cm {\large
{\bf Francesco Coradeschi}$^a$, 
{\bf Paolo Lodone}$^a$, {\bf Duccio Pappadopulo}$^{b,c}$, \\
{\bf Riccardo Rattazzi}$^a$
and {\bf Lorenzo Vitale}$^a$ 
} \\[1cm]
{\it 
$^a$ Institut de Th\'eorie des Ph\'enom\`enes Physiques, EPFL, Lausanne, Switzerland \\
$^b$ Department of Physics, University of California, Berkeley, CA 94720, USA\\
$^c$ Theoretical Physics Group, Lawrence Berkeley National Laboratory, Berkeley, USA
} \\[5mm]
\vskip 1.0cm
\end{center}

\begin{abstract}\noindent
Goldstone's theorem does not apply straightforwardly to the case of spontaneously broken scale invariance.
We elucidate under what conditions a light scalar degree of freedom, identifiable with the dilaton, can naturally arise.
Our construction can be considered  an explicit dynamical solution to the cosmological constant problem in the scalar version of gravity.

\end{abstract}
\end{titlepage}


\section{Introduction}

The purpose of this paper is to discuss the conditions under which a light scalar, identifiable with the dilaton, can naturally arise in a field theory \cite{cpr}. This question is non-trivial because dilatation invariance is a spacetime symmetry, and  Goldstone theorem does not apply straightforwardly. To put the problem into focus, let us then review the basic facts.

 In the case of a non-linearly realized ordinary global symmetry, the Goldstone field transforms by a simple constant shift:
\beq
\tau(x)\to \tau(x)+c
\eeq
so that the only scalar potential consistent with the symmetry vanishes identically $V(\tau)\equiv 0$. Then, not only the mass but all interactions vanish at zero external momentum. In the case of dilatation invariance the associate Goldstone scalar transforms instead as:
\beq
\tau(x)\to \tau(kx)+\ln k
\eeq
with $k\in {\mathbb R}^+$. Consistent with dilatation invariance the most general scalar potential is then:
\beq \label{eq1:PotentialForPi}
V=V_0e^{4\tau}
\eeq
with $V_0$ a generically non-vanishing constant with dimension $[E]^4$. This state of things implies that the pattern of symmetry breaking depends on the parameter $V_0$, as we shall now illustrate.

Aside from the potential, the  most general dilatation-invariant Lagrangian for $\tau$ will include higher derivative terms
with the schematic form
\beq
e^{4\tau} \left (\partial\right )^m \left (e^{-\tau}\right )^m
\eeq
and with the $m$ partial derivatives spread over the $e^{-\tau}$ factors in all possible ways.
Notice that, while the potential (\ref{eq1:PotentialForPi}) is also invariant under the full conformal group $O(4,2)$, only very specific combinations of the higher derivative terms are invariant under special conformal transformations.
To be specific, the most general conformal-invariant action  $\mathcal{S}_{CI}[\tau]$ can be constructed as the most general diffeomorphism-invariant action  involving the metric \cite{Salam:1970qk}:
\beq
\hat{g}_{\mu\nu} = e^{2\tau} \eta_{\mu\nu}
\eeq
plus a single ``Wess-Zumino'' term that cannot be written in this form\cite{Tomboulis:1988gw,Nicolis:2008in}:
\beq \label{eq:CILagrangian}
\mathcal{S}_{CI}[\tau] = \mathcal{S}[\hat{g}] + \mathcal{S}_{WZ}[\tau] \, .
\eeq
For simplicity from now on we assume the case (\ref{eq:CILagrangian}), with invariance under the full conformal group.
It would be perhaps interesting to study whether there can be substantial changes in our discussion in the case of scale-without-conformal invariance. 

The presence of an explicit dimensionful parameter, $V_0$, is just due to our use of a dimensionless field $\tau$ and is obviously  consistent with a non-linearly realized dilatation invariance. To make the symmetry more evident we will also work with a canonical dilaton field $\varphi\equiv f_De^\tau$, in terms of which the most general effective Lagrangian truncated at two derivatives 
is:
\beq \label{eq:withKappa}
{\cal L}=\frac{1}{2}\partial_\mu\varphi\partial^\mu\varphi -\kappa\varphi^4
\eeq
 with $\kappa=V_0/f_D^4$ a dimensionless coupling. Focussing on this simplest Lagrangian, as already mentioned,  the pattern of symmetry breaking depends on the parameter $\kappa$. That can be studied  by considering the maximally symmetric solutions, as first done in ref. \cite{fubini}. One finds:
\begin{eqnarray}\label{pattern}
\kappa&>&0 \quad\to\quad \varphi=\frac{1}{\sqrt{2\kappa}}\frac{1}{z}\qquad SO(3,2)\equiv {\rm AdS4}\nonumber \\
\kappa&=&0 \quad\to\quad \varphi={\rm const}\qquad ISO(3,1)\equiv {\rm Poincar\acute{e}\, 4}\\
\kappa&<&0 \quad\to\quad \varphi=\frac{1}{\sqrt{-2\kappa}}\frac{1}{t}\qquad SO(4,1)\equiv {\rm dS4} \, .\nonumber
\end{eqnarray}
 As a matter of fact the result does not qualitatively change when considering the most general conformally invariant derivative action  \cite{Nicolis:2008in,Nicolis:2009qm}. It then follows that the spontaneous breakdown of $O(4,2)$ to Poincar\'e, with a resulting massless dilaton, does not arise for a generic choice of parameters, but requires 
 the tuning $\kappa=0$. As far as we know the only case in which  the choice $\kappa=0$ is technically natural is in the context of supersymmetry.
 There, in particular in $N=4$ Super Yang-Mills, there are plenty of flat directions that can play the role of the dilaton.
Notice also that eq.~(\ref{pattern}) corresponds precisely to the situation in general relativity: depending on the sign of the cosmological constant $\Lambda$ there are either {\it dS} or {\it AdS} solutions,  while only for the special choice $\Lambda =0$ is the solution Poincar\'e invariant. This is not surprising given that the action for the conformal mode of the metric and that of the dilaton share invariance under  $O(4,2)$. In this respect the tuning associated with a massless dilaton is completely analogous to the tuning associated with  a vanishing cosmological constant in gravity \cite{Sundrum:2003yt}. A solution of the former problem may hopefully shed light on the latter.

The breaking pattern \ref{pattern} resembles that of the Lorentz group $SO(3,1)$ when considering  a vector field $A_\mu$ with a potential 
\beq
V \propto (A_\mu A^\mu -m^2)^2 \, .
\eeq
Depending on $m^2$, the minimum  is in fact at $A_\mu = \hat{A}_\mu$, where $\hat{A}$ can be chosen to be:
\begin{eqnarray}\label{patternLorentz}
m^2&<&0 \quad\to\quad \hat{A} = (0,|m|,0,0) \qquad SO(2,1) \nonumber \\
m^2&=&0 \quad\to\quad \hat{A} = (p,p,0,0) \qquad ISO(2)\\
m^2&>&0 \quad\to\quad \hat{A} = (m,0,0,0) \qquad SO(3) \, .\nonumber
\end{eqnarray}
Notice also the analogy with the theory of representations of the Lorentz group, where the residual symmetry group in (\ref{patternLorentz}) is the little group and $m^2$ is the squared momentum of a one-particle state.
Then in the case of representations with spin $1/2$ or $1$ the massless case can be selected  respectively by chiral or gauge symmetry, or, more generally, by multiplet shortening. On the contrary for spin $0$ it is unnatural to have $m^2=0$,
also related to the absence of multiplet shortening at $m=0$. This is the source of the well-known hierarchy problem.

These  simple examples illustrate that the non-compact nature of the group ($O(4,2)$ or $SO(3,1)$) plays a central role to produce a ``phase diagram" where some specific  breaking pattern (to $ISO$ groups) can arise only on a subspace of zero measure, that is by tuning.
Indeed if we considered the same vector $A_\mu$ but with compact symmetry group $SO(4)$ the breaking pattern would more simply  be 
\begin{eqnarray}\label{patternEuclid}
m^2&\leq&0 \quad\to\quad \hat{A} = (0,0,0,0) \qquad SO(4) \nonumber \\
m^2&>&0 \quad\to\quad \hat{A} = (m,0,0,0) \qquad SO(3) \, .\nonumber
\end{eqnarray}
so that the breaking pattern presents only two, generic, options.

In phenomenological applications we are often interested in pseudo-Goldstone bosons, whose mass results from the explicit  breaking of the global symmetry by a small parameter. In the case of internal compact symmetries, the possible symmetry breaking patterns are robust and generic, as seen in the $SO(4)$ example mentioned above. It is thus straightforward to apply an explict symmetry breaking perturbation, the pion in QCD being a perfect example. On the other hand,  our discussion  shows that for
dilatations  the very starting point is non-generic and  seemingly implausible. Further  elaboration is thus needed to identify  a naturally light dilaton.

The discussion so far concerned the case of exact conformal symmetry.  The next obvious step is to ask what happens in the presence of a (small) explicit source of breaking.
Consider now the case of explicit breakdown of conformal invariance, where couplings $\lambda_i$ that take the system away from the fixed point are turned on:
\beq
\mu\frac{\partial \lambda_i}{\partial \mu}=\beta_i(\lambda)\not = 0\, .
\eeq
 The simplest  and perhaps most interesting case is that of just 
one relevant or marginally relevant  coupling $\lambda$ associated with:
 \beq
 \Delta {\cal L} = \lambda {\cal O}_{d}
 \eeq
where ${\cal O}_{d}$ has dimension  $d\leq 4$ in the limit  $\lambda=0$. In this situation, by starting  at some UV scale $\mu_0$ with  $\lambda(\mu_0)\equiv \lambda_0\ll 1$,
 the system is driven further away from the fixed point by the Renormalization Group (RG) flow towards the IR, until at some  scale $\Lambda$ one has $\lambda(\Lambda)\sim1 $ corresponding to a $O(1)$ perturbation\footnote{We apply Naive Dimensional Analysis (NDA)  normalizing the couplings so that the perturbation expansion parameter is $\lambda$ without extra powers of $4 \pi$.} away from conformality. The resulting physics is strongly coupled and generically  characterized by just one scale $\Lambda$, like in QCD, with masses scaling in units of $\Lambda$. Massless, or light degrees of freedom, will be associated with broken global symmetries (Goldstone bosons) or with unbroken chiral and gauge symmetries (respectively fermions and vector bosons). However, since  conformal invariance is no longer an approximate symmetry at the relevant energy scale, witness the fact that the coupling $\lambda$ runs `fast',  there is no reason to expect a light dilaton-like CP even scalar.
More explicitly, this is because the non-conservation of the scale current $S_\mu$ is controlled by the beta function:
\beq
\partial^\mu S_\mu = T_\mu^\mu \propto \beta(\lambda) \, .
\eeq
The one we outlined is indeed the situation realized in UV free gauge theories like QCD, with the NDA normalized gauge coupling $g/4\pi$ playing the role of $\lambda$. Expectedly there is no candidate light and narrow dilaton in the observed hadron spectrum. 
Similarly no light  dilaton was to be expected in ordinary technicolor models. Moreover
in conformal technicolor models like the one proposed in \cite{Evans:2010bp}, where the role of $\lambda$ is played by a very relevant coupling such as  a fermion mass,
we do not expect a light dilaton-like state. Again this is because, at the relevant IR scale $\Lambda$, conformal invariance is not anymore an approximate symmetry. Notice that this situation does not change at large $N$.

In the following Sections we shall illustrate under what conditions this generic expectation fails and a naturally light dilaton-like scalar emerges. More precisely, in  Section \ref{sec:4Dpicture} we illustrate the requirements from a purely 4-dimensional point of view. 
In Section \ref{sec:hologrealiz} we discuss a 5D model representing an explicit  holographic realization. This construction allows to perhaps better evaluate the plausibility  of the requirements sketched in the purely 4-dimensional discussion.
In Section \ref{sec:conclusions} we briefly draw our conclusions. Various additional aspects of the 5D model, such as the stability of the solution, are discussed in the appendices.

\section{The 4-dimensional picture} \label{sec:4Dpicture}

First of all we should make clear that, underlying our all discussion is the assumption that we are dealing with a CFT where some non trivial operators acquire non-vanishing expectation value thus spontaneously breaking dilatations. Under this assumption, we will focus from now on on the effective theory of the resulting dilaton, addressing the problem that was outlined in the Introduction.

The discussion in the Introduction, in spite of being negative, does suggest the  features that are necessary in order to obtain a naturally light dilaton. A we shall now elaborate, these are:
\begin{enumerate}
\item
The CFT should somehow be able to sample a direction with $\kappa=0$ in (\ref{eq:withKappa}).
\item It should be endowed with a coupling that stays `naturally' close to marginality throughout the RG evolution.
\end{enumerate}
The first request can be satisfied by postulating that the theory possesses a line (or more generally a surface) of fixed points.
This corresponds to the existence of a coupling $\lambda$ (or a set of them) that remains exactly marginal over a finite range. The corresponding marginality line (or surface) can be viewed as a continuous family of CFTs that are deformed into one another by turning on the exactly marginal coupling. Now, the parameter $\kappa$ will vary continuously over this family, $\kappa\to \kappa(\lambda)$, and generically there will exist a point $\lambda_*$, or a discrete set, such that  $\kappa(\lambda_*)=0$. To satisfy the second request,  imagine now to modify the theory by endowing $\lambda$ with a small beta function over the whole marginality line:
\beq
\beta(\lambda)=\epsilon \bar\beta(\lambda)\qquad \epsilon \ll 1\, , \quad \bar\beta(\lambda)=O(1)\, . \eeq
By RG invariance the dilaton potential will simply be\footnote{We imagine that $\epsilon$ smoothly describes a one parameter family of theories and work
in series expansion in $\epsilon$ around $\epsilon = 0$.  Eq. (\ref{eq:potentialKappaPhi4}) represents the potential at zeroth order in $\epsilon$. The holographic example we shall present later supports this picture.
Higher order effects will modify the function $\kappa$, but its relevant properties, zeroes and slope, will qualitatively remain the same over a finite range of $\epsilon$. So we can neglect this detail in the discussion.}:
\beq \label{eq:potentialKappaPhi4}
V(\varphi)=\kappa(\lambda(\varphi)) \varphi^4\, .
\eeq
This basically corresponds to a quartic potential modulated by a slow evolution with $\varphi$ of its coefficient $\kappa$, the slow dependence arising from the near marginality of $\lambda$. Now, by a generic choice of parameters, one that does not require any particular tuning, we can imagine $\kappa(\lambda(\varphi))$ to be positive at $\varphi\to \infty$ and to cross zero at $\varphi=\varphi_*$ such that $\lambda(\varphi_*)=\lambda_*$. In such situation  the minimum of the potential will clearly be at $\varphi=O(\varphi_*)$, close to the point where the quartic coefficient vanishes. The resulting mass of the dilaton will thus be suppressed by $\epsilon$, the small parameter in the game.
This result is precisely what happens in dimensional transmutation \`a la Coleman-Weinberg \cite{Coleman:1973jx}.
To make the discussion more quantitative, we can study the vacuum dynamics  in an expansion in $\epsilon$ around $\varphi_*$ ($\lambda(\varphi_*)=\lambda_*$ and $\kappa(\lambda_*)=0$).
The condition of stationarity 
\beq
\frac{\partial V(\varphi)}{\partial\varphi}=\left [4\kappa(\lambda(\varphi))+\beta(\lambda(\varphi))\kappa^\prime(\lambda(\varphi))\right ]\varphi^3=\left [4\kappa(\lambda(\varphi))+\epsilon \bar\beta(\lambda(\varphi))\kappa^\prime(\lambda(\varphi))\right ]\varphi^3\eeq
implies the minimum is at a $\varphi_{min}$ satisfying:
\beq
\lambda(\varphi_{min})\equiv \lambda_{min}=\lambda_*-\frac{\epsilon}{4} \bar\beta(\lambda_*) +O(\epsilon^2)
\eeq
implying
\beq
\varphi_{min}=\varphi_*e^{-\frac{1}{4}+O(\epsilon)}\, .
\eeq
Assuming, without loss of generality, a canonically normalized kinetic term, we find for the dilaton mass:
\beq \label{eq:mphiGenericDiscussion}
m^2_\varphi=4\epsilon \varphi_{min}^{2} \bar\beta(\lambda_*) \kappa^\prime(\lambda_*)=O(\epsilon)\varphi_{min}^2
\eeq
suppressed with respect to the characteristic mass scale of the system $\varphi_{min}$.

\section{Holographic realization}   \label{sec:hologrealiz}

To better appreciate how plausible the scenario of the previous Section is,  we outline here a holographic realization \cite{Maldacena:1997re}-\cite{Verlinde:1999fy}, in the context of RSI \cite{Randall:1999ee}.
Our mechanism is a variant of the one proposed in \cite{Goldberger:1999uk} by Goldberger and Wise (GW). We do not want to claim particular originality here: see \cite{Goldberger:1999uk}-\cite{Eshel:2011wz} for related studies and \cite{Chacko:2012sy}\cite{Bellazzini:2012vz} for recent discussions\footnote{During the completion of this work, another paper appeared \cite{Bellazzini:2013fga} in which the idea of \cite{cpr} is elaborated.}. 
We just want to elucidate in the holographic context the necessary conditions for a naturally light dilaton, that here coincides with the so-called radion.  In that respect our remarks complement the discussion in the Appendix A of ref. \cite{Rattazzi:2000hs}.

We want to translate into an AdS5 model the properties of the CFT we previously identified. A naturally marginal deformation will correspond to a naturally massless scalar in 5D, a Goldstone boson $\pi$ living in the bulk:
\beq \label{eq:duality0}
\pi\quad \leftrightarrow \quad\lambda \, .
\eeq
The marginality surface in the CFT  will correspond to the coset manifold in AdS5.
 An almost marginal deformation, like the one we want, will then just correspond to a bulk pseudo-Goldstone. We will parametrize with a small dimensionless quantity $\epsilon$ the effects that explicitly break the Goldstone symmetry. In particular the bulk scalar potential $V(\pi)$ will be $O(\epsilon)$. 
The tension $\tau_{IR}$ of the IR brane contributes additively to the dilaton quartic $\kappa$ \cite{Rattazzi:2000hs}. Then the request 1) in the previous Section amounts to assuming a $\pi$-dependent tension. 
We choose units where the AdS5 radius is $L=1/k$, and the bulk cosmological constant is $\Lambda_5=-3/L^2$. 
 For the infrared (IR) brane tension we shall assume (see eq.~(\ref{eq:fullaction0}) for the units) :
\beq \label{eq:IRtension}
\tau_{IR}(\pi)=-\frac{3}{L} +\frac{f(\pi)}{L} =\tau_{RS}+\frac{f(\pi)}{L} 
\eeq
where $\tau_{RS}$ is the tuned value corresponding to an exactly vanishing dilaton potential, that is $f=0$ corresponds to $\kappa=0$. So we basically have:
\beq
\kappa\equiv \kappa(\lambda) \quad \leftrightarrow \quad 
{f(\pi)} \, .
\eeq
In our study we shall  elucidate  the relation between $\kappa(\lambda)$ and $f(\pi)$. While the relation will become conceptually clear, we shall only be able  to present simple analytic expressions  under some approximations: a first order Taylor expansion in the case of large back-reaction  from the field $\pi$ in Sections \ref{subsec:radion}-\ref{subsec:radion2}, to all orders in $\lambda$ for the case of small back-reaction  in Section \ref{subsec:radion0}.

We now look for a solution with a  5D metric 
\begin{equation} \label{eq:metric}
ds^2 = g_{NM} dx^N dx^M= e^{2 A(z)} \eta_{\mu\nu} dx^\mu dx^\nu - dz^2
\end{equation}
where $\eta_{\mu\nu}=\mbox{diag}(1,-1,-1,-1)$.
Notice that (\ref{eq:metric}) is the most general Poincar\'e invariant solution, after the gauge choice $g_{\mu 5}=0$ and $g_{55}=-1$.
To allow for a holographic interpretation we focus on asymptotically AdS solutions, i.e. we impose $A(z) \rightarrow -z/L$ at $z\to -\infty$.
We introduce an IR brane at $z=z_{IR}$, whose presence is associated with the spontaneous breakdown of 4D conformal invariance \cite{ArkaniHamed:2000ds}\cite{Rattazzi:2000hs}, and for simplicity we do not introduce any ultraviolet (UV) brane. This corresponds to the limit of zero Newton constant in 4D, which is legitimate since our considerations are intrinsecally decoupled from 4D gravity. Introducing the Planck brane one would slightly complicate the discussion and find the usual issue of the finetuning related to the 4D Cosmological Constant.
Adopting the conventions of \cite{DeWolfe:1999cp}, our 5D action is thus given by:
\begin{equation} \label{eq:fullaction0}
\frac{S}{(M_5)^3}=\int d^4 x \int_{-\infty}^0 dz \sqrt{|g|} \left[ -\frac{1}{4} R + \frac{1}{2}(\p \pi)^2 - V(\pi)\right ]
-\frac{1}{2} \int_{z=z_{IR}} d^4 x \sqrt{|h|} \left[ \tau_{IR}(\pi)  + K \right]
\end{equation}
where $h$ is the 4D metric induced on the brane, and $K$ is the extrinsic curvature\footnote{Notice that performing a 4+1 split and using ADM variables \cite{Arnowitt}, as done for example in \cite{Luty:2003vm}, the ``Gibbons-Hawking'' term involving $K$ is automatically canceled and only first derivatives in $z$ appear (see also \cite{Gibbons:1976ue} \cite{Wald:1984rg}).} of the boundary (brane).
Notice that the 5D Planck scale $M_5$ is factored out and thus will not enter  our considerations. 
We will parameterize our potential by
\begin{equation} \label{eq:PhiPotential}
V(\pi) = -\frac{3}{L^2} +\frac{\epsilon}{L^2} P(\pi ) \, ,
\end{equation}
where $\epsilon$ is a small parameter controlling the explicit breaking of the Goldstone symmetry $\pi \to \pi + c$. Notice that, while the shift symmetry is broken  by the small parameter $\epsilon$ in the bulk, it is instead maximally broken by the tension at the IR boundary, see eq.~(\ref{eq:IRtension}). This situation is technically natural because of the locality of the UV divergent corrections to the $\pi$ potential: the breakdown of the Goldstone symmetry at the boundary cannot affect the bulk potential \footnote{We expect finite quantum effects  to asymptotically vanish   away from the IR brane as $z\to -\infty$.}.
According to the AdS/CFT dictionary (see for instance ref.~\cite{Girardello:1998pd}), the dual running coupling $\lambda$ can be identified with $\pi $, while the corresponding $\beta$ function is 
\beq
\beta(\lambda) =\frac{\epsilon}{4}\partial_\lambda P(\lambda)\equiv \epsilon \bar\beta(\lambda)\, .\label{betadual}
\eeq
By a  direct inspection of the equations of motion, the condition to have  an asymptotic AdS space at $z\to -\infty$,
 (that is   for the   $\pi$ field back reaction on the metric  to vanish asymptotically) is
 \beq
\epsilon P''(0) < 0\, .\label{UVfreedom}
\eeq
This condition precisely corresponds  to the UV stability of the unperturbed $\lambda = 0$ fixed point in the dual CFT description.
In what follows we shall assume $\epsilon >0$, $P''(0)<0$ without loss of generality. We shall also present some more details on the simple case of a quadratic tachyonic potential
$P=-2\pi ^2$, for which $\beta = -\epsilon \lambda$, corresponding to a perturbation with fixed scaling dimension $\epsilon$. However our discussion applies to a generic 
flat potential ($\epsilon \ll 1$).
%

Consider now the equations of motion (EOM) that come from the variation of (\ref{eq:fullaction0}) when both $A$ and $\pi$ are functions of $z$ only, which corresponds to the most general solution with Poincar\'e symmetry. In the bulk the EOM read:
\begin{eqnarray}
\pi''+4 A' \pi' - \frac{\p V}{\p \pi} &=& 0 \label{eq:EOM1}\\
 A''+ \frac{2}{3}(\pi')^2&=&0 \label{eq:EOM2} \\
(A')^2+ \frac{1}{3}V(\pi) -\frac{1}{6}(\pi')^2&=&0 \, , \label{eq:EOM3}
\end{eqnarray}
where here and below the primes denote the derivatives with respect to $z$, supplemented by the matching conditions on the brane:
\begin{eqnarray}
A' (z=z_{IR}) &=& \frac{1}{3} \tau_{IR}(\pi(z_{IR})) \label{eq:Matching1} \\
\pi' (z=z_{IR}) &=& -\frac{1}{2} \frac{\p \tau_{IR} (\pi(z_{IR}))}{\p \pi} \label{eq:Matching2} \, .
\end{eqnarray}
For $\epsilon=0$ these equations can be solved exactly and the solution is given by:
\begin{eqnarray}
A_0(z) = \frac{1}{4}\log \sinh  \frac{4(z_*-z)}{L}  - \frac{z_*}{L} + \frac{\log 2}{4} && [\epsilon=0] \label{eq:ASolutionZero} \\
\pi_0(z) = \pm \frac{\sqrt{6}}{4} \log \tanh \frac{2(z_*-z)}{L}   + \pi_*  && [\epsilon=0]\, . \label{eq:PhiSolutionZero} 
\end{eqnarray}
The additive constant in $A(z)$ is fixed by the boundary condition $A(z)\to -z/L$ at $z\to -\infty$. The integration constants $z_*$ and $\pi_*$ are instead determined by the matching conditions (\ref{eq:Matching1})-(\ref{eq:Matching2}) once the tension is specified as a function of $\pi$. More precisely one has 
\beq
z_*=z_{IR}+ c_* L\label{zstar}\, .
\eeq
where, assuming $\tau_{IR}(\pi)$ is a generic $O(1)$ function, we expect $c_*$ to be of order 1. Moreover,  $c_*$ must be positive, otherwise the solution has a singularity at $z<z_{IR}$. In general there is only a discrete set of solutions and thus, up to a discrete ambiguity that is not important for our discussion, the parameters $c_*$ and $\pi_*$ are fixed by the dynamics, that is by $\tau_{IR}(\pi)$. Notice in particular that $c_*$ and $\pi_*$  do not depend on $z_{IR}$: by varying $z_{IR}$, and $z_*$ according to eq.~(\ref{zstar}), we obtain a family of solutions, satisfying the same boundary conditions.  We conclude that $z_{IR}$ is a modulus and that the associated 4D scalar field in the Kaluza-Klein decomposition, the radion, must be massless. The presence of this modulus suggests we must have made a tuning. Where? Notice indeed  that
we did not fix a priory the boundary condition of $\pi$ at $z\to -\infty$, but rather determined it from the very existence of a Poincar\'e invariant solution. From eq.~(\ref{eq:PhiSolutionZero}) one finds $\lim_{z\to -\infty}\pi(z)=\pi_*$, a parameter purely fixed by the IR boundary condition. For any other  choice of the  asymptotic value of  $\pi$  there would  not exist a solution with Poincar\'e invariance. For these other choices we 
should  find solutions with either dS or AdS residual isometry.

The holographic dual of the above state of affairs is precisely what we described in Section \ref{eq:duality0}: only for the specific choice $\lambda = \lambda_*$ of the marginal coupling do we have a vanishing dilaton potential allowing the breaking of $O(4,2)$ to the  Poincar\'e group. For all other choices the breaking is  either to dS or to AdS.
As already said, this fine-tuning, corresponding to $\kappa\rightarrow 0$ in (\ref{eq:withKappa}), is an analogue of the Cosmological Constant problem in the scalar version of gravity \cite{Sundrum:2003yt}. We will see in the following how our construction can be considered as a dynamical solution to this problem.

As we already remarked, $z_{IR}$ is a modulus. The corresponding family of solutions can simply be obtained
 by performing the (global) change of coordinates
\beq
z\to \tilde z= z-z_1\qquad\qquad x^\mu \to \tilde x^\mu =x^\mu e^{-z_1/ L}\, .
\label{coordinateshift}
\eeq
which leaves the UV boundary conditions $A(z) \to -z/L$, $\pi \to \pi _*$ at $z\to -\infty$ unaffected, and which in practice just amounts to the shift
\bea
z_{IR} &\rightarrow & z_{IR} - z_1 \equiv \tilde z_{IR}  \nonumber\\
z_* &\rightarrow & z_* - z_1 \equiv \tilde z_*\, .\label{eq:shiftTransformation}
\eea
By this result the action is stationary under variations of $z_{IR}$, consistent with it being a modulus.
Notice that eq.~(\ref{coordinateshift}) precisely corresponds to a 4D dilatation in the dual picture. Under this change of coordinates, the warp factor at the IR boundary changes
as
\beq
e^{A_{IR}}\equiv e^{A(z_{IR})} \to e^{\tilde A_{IR}}= e^{A_{IR}} e^{z_1/L}\, .
\eeq
That is precisely how  the dilaton $\varphi\propto e^\tau$ transforms. This is consistent with the familiar result from  RS phenomenology, where  the warp factor at the IR boundary  can be interpreted, up to an overall normalization, as the interpolating field for the canonical dilaton $\varphi$.

In Section \ref{subsec:radion} we  shall discuss the dilaton mode in more detail. Moreover, in order to assess the validity of the solution we just found, we should also insure that it is  stable, i.e that   no Kaluza-Klein mode around it is a ghost or a tachyon. It can easily be checked  that there are no ghosts, while in  Appendix \ref{appendix:KK} we prove  that tachyons are avoided by a mild and generic request on the IR brane tension: ${\p^2 \tau_{IR} (\pi(z_{IR}))}/{\p \pi^2}>0$.

\subsection{The 5D solution at $\epsilon\not = 0$} \label{sec:5DsolutionEpdZero}

Let us consider now the case where $\epsilon \not = 0$. In general the equations of motion cannot be solved exactly.
In principle we could imagine to proceed by treating $\epsilon $ as a small perturbation and  by expanding the solution in a power series
\bea
\pi&=&\pi_0+\epsilon \pi_1+\epsilon^2 \pi_2 +\dots\label{piseries}\\
A'&=& A'_0 +\epsilon A'_1+\epsilon^2 A'_2 +\dots\label{Aseries}
\eea
where $\pi_0$ and $A_0'$ are the zeroth order solutions in eqs.~(\ref{eq:ASolutionZero})-(\ref{eq:PhiSolutionZero})
However that only works for finite $z$. To see more explicitly what happens, let us  fix (withouth loss of generality), $z_*=0$ in the unperturbed solution. We thus have $z_{IR}=-c_* L= O(L)$. We can then solve the equations of motion, order by order in $\epsilon$ starting from the IR brane: eqs.~(\ref{eq:EOM3})-(\ref{eq:Matching2}) fix the initial conditions for $\pi$, $\pi'$ and $A'$ and  the solution ($\pi,A'$) is unique. The warp factor $A$ is then obtained by performing a further integration:  the overall additive constant can for instance be fixed by the request $\lim_{z\to -\infty}A(z) = -z/L$. Now, notice that the unperturbed solution quickly enters its asymptotic behaviour at  $-z/L > O(1)$
\beq
\pi_0(z) =\pi_*+ O(e^{-4|z|/L})\qquad\qquad A'_0(z)= -\frac{1}{L} +O(e^{-8|z|/L})\, .\label{asymptotes}
\eeq
Using this result, by studying the linearized second order differential equation for $\pi_1$,  in the region $-z/L\gg 1$  one finds
\beq
\epsilon \pi_1 = - \frac{\epsilon z}{4L} P'(\pi_*)+O(e^{-4|z|/L})\, .\label{firstorder}
\eeq
 We conclude that for generic $P$ (in particular for quadratic $P$) we can treat the potential as a perturbation only as long as $\epsilon|z|/L \ll 1$. Moreover, in the region $ \ln 1/\epsilon \ll |z|/L \ll 1/\epsilon$, the exponentially decaying part in $\pi_0$ and $A'_0$ is subdominant to the $O(\epsilon)$ perturbation. In that region,    to first non-trivial order, the solution is then
 \beq
 \pi=\pi_*- \frac{\epsilon z}{4L} P'(\pi_*)+O(\epsilon^2)\qquad\qquad A'(z)= -\frac{1}{L} +\frac{\epsilon}{L}P(\pi_*)+O(\epsilon^2)\label{nontrivial}
 \eeq
The above equations provide the initial matching conditions in the region $ \ln 1/\epsilon \ll |z|/L \ll 1/\epsilon$ for the solution at large $z$. Indeed, by inspecting the equations of motion, one readily concludes that the solution  matching eq.~(\ref{nontrivial}) satisfies, to leading order in $\epsilon$, a first order differential equation
\beq
\pi' = \frac{-\epsilon }{4L} P'(\pi) \qquad \qquad A'(z)= -\frac{1}{L} +\frac{\epsilon}{L}P(\pi)
\label{ansatz}
\eeq
with (at leading order in $\epsilon$) boundary condition $\pi=\pi_*$ at $|z|/L=O(1)$. 
Indeed eq.~(\ref{ansatz}) is consistent with eq.~(\ref{nontrivial}) in the matching region, and when substituted into the equations of motions  solves them up to $O(\epsilon^2)$ terms. In particular the term $\pi''$ in eq.~(\ref{eq:EOM1}) is of order $\epsilon^2$
according to eq.~(\ref{ansatz}), and thus subleading. The evolution of $\pi$ towards the conformal boundary thus follows
a first order differential equation, whose CFT interpretation is the  RG equation for the dual coupling . The solution of eq.~(\ref{ansatz}) amounts to a resummation of all powers of $\epsilon z /L$ as $z\to -\infty$, while the neglected terms correspond to next-to-leading order powers $\epsilon (\epsilon z/L)^n$. The analogy with the RG resummation of leading logs is obvious.

Notice that we worked under the assumption of asymptotic AdS geometry at $z \to - \infty$. It is therefore essential, for our whole picture to make sense, that the contribution of $\pi$  to the energy momentum tensor vanish  towards the boundary.  
 A sufficient condition for this to happen is that\footnote{Notice that $P(0)=0$ can always be achieved by redefining the bulk cosmological constant, while a stationary point  $P'=0$ can always be set at $\pi=0$ by redefining $\pi$ via a constant shift.} $P(0)=0, P'(0)=0$ and $P''(0)<0$, in which case $\lim_{z\to -\infty}\pi = 0$
 for some finite range of  $\pi_*$. This situation
corresponds to a UV stable fixed point in the dual theory. An example satisfying this criterion is given by
the quadratic potential $P= -2\pi^2$.
In this case the form of the solution in the asymptotic region is
\begin{equation} \label{eq:AsymptPhiSolution}
\pi(z \ll -L/\epsilon) = \pi_* e^{\epsilon z/L}\, \left( 1 + O(\epsilon e^{2\epsilon z/L}) \right) + \hat{\pi}_{*} e^{(4-\epsilon)z/L} \, \left( 1 + O(\epsilon e^{2\epsilon z/L})\right)  \, .
\end{equation}
where  $\hat{\pi}_*$ is an  ``integration constant'' determined by the matching conditions including subleading terms, which we disregarded in the above general  discussion. The leading term, scaling like $e^{\epsilon z/L}$, precisely corresponds to the solution of eq.~(\ref{ansatz}).
Integrating $A'$ from eq.~(\ref{eq:EOM3})  using (\ref{eq:AsymptPhiSolution}),  we find  the leading  correction to the AdS behaviour of the metric 
\begin{equation} \label{eq:AsymptASolution}
A(z \ll -L/\epsilon) = -\frac{z}{L} +\frac{\pi_*^2 }{3}   \left( 1-e^{2\epsilon z/L}+ O(\epsilon e^{2\epsilon z/L}) \right)  \, .
\end{equation}
Notice  that the additive constant $\frac{ \pi_*^2 }{3}$ can in principle be removed  in order to satisfy
the boundary condition $\lim_{z\to -\infty}A(z) = -z/L$. However, if that is done, then in the region near the IR brane $A-A_0=O(1)$, while $A'-A_0'=O(\epsilon)$ everywhere.


Our solution of the $\epsilon \not = 0$ case was obtained by perturbing around a given choice of the IR brane coordinate $z_{IR}=-c_* L$, corresponding to $z_*=0$ in the unperturbed solution. It is pretty obvious that the asymptotic behaviour of $\pi$, which is now not constant, will depend on this choice. This is seen clearly by performing the coordinate shift in eq.~(\ref{coordinateshift})  which does not affect the asymptotic behaviour of the warp factor $A(z)$ but does change the asymptotic behaviour of $\pi$
\beq
z_{IR}\to z_{IR}-z_1\qquad\qquad \pi(z)\to \pi(z+z_1)\, .\label{RG}
\eeq
According to this equation the position of the IR brane is in one to one correspondence with the value of the ``running" field at any given test scale $z$. This is seen explicitly in the case of a quadratic potential, for which the shift in the solution can be translated into a change of its overall coefficient  in eq.~(\ref{eq:AsymptPhiSolution})
\beq \label{eq:shiftTransformation2}
\pi_* \rightarrow \pi_* e^{ \epsilon z_1/L} \, .
\eeq
Notice the change with respect to the $\epsilon =0$ case. In that case, $\pi$ evolves to an undetermined constant at $z\to -\infty$: in order to obtain a Poincar\'e invariant solution, we must  tune the constant to be equal to $\pi_*$. Moreover the leading behaviour at infinity is not affected by a shift of the IR boundary, so we expect the radion to be exactly massless. In the case $\epsilon \not =0$, the Poincar\'e invariant solution is generic. Now the field $\pi$ is automatically driven to a fixed point $\pi=0$ at $z\to -\infty$, 
so that now the boundary condition on $\pi$ must specify
the rate at which it approaches $0$. One convenient prescription  is to pick a fixed value $\pi_{UV}$ near zero and define the boundary condition in terms of the value $z_{UV}$ of $z$ such that $\pi(z_{UV}) = \pi_{UV}$. Keeping $\pi_{UV}$ fixed and shifting $z_{UV}$, by  eq.~(\ref{RG}) 
$z_{IR}$ shifts by the same amount. This correlation between the location of the IR brane and the choice of boundary condition, implies the radion is stabilized. Moreover, in the limit where $\epsilon$ is small the radion mass will obviously be small.
 By translating  to the 4D dual picture via the dictionary
\beq
\pi(z) \to \lambda(\mu) \qquad\qquad \frac{e^{-z/L}}{L} \to \mu \qquad\qquad  \frac{e^{-z_{IR}/L}}{L}\to \langle \varphi\rangle
\eeq
one finds agreement with the discussion in Section  \ref{sec:4Dpicture}.
Our 5D model can be thus considered a dynamical solution to the Cosmological Constant problem of scalar gravity \cite{Sundrum:2003yt}.

%



Our logic  implies, for $\epsilon \ll 1$, a radion with mass $m_\varphi$ parametrically smaller than the Kaluza-Klein gap $m_{KK}\sim e^{-z_{IR}}/L$, corresponding to an effectively small quartic $\kappa$ around the minimum. In the remaining Sections of the paper we shall compute the dilaton effective action
working at the first non-trivial order in $\epsilon$. We shall proceed in three steps, as follows.

\noindent{\bf  1.} We shall first find the radion/dilaton effective Lagrangian at $\epsilon =0$ and truncated to two derivative terms. In practice we shall find a mode that acts as a good interpolating field for the dilaton. In particular, at zero momentum it is diffeomorphic to the unperturbed solution, implying that its action involves at least two derivatives. As shown in Appendix A, that property also holds true when    KK excitations are turned on.  The two derivative effective action is then simply obtained by substituting the mode into the 5D action, while the effect of massive KK exchange affects the action starting at four derivatives. This is because the mixing between the radion and any massive KK mode starts at  $O(\partial^2)$: integrating out the KK at tree level one obtains a $O(\partial^4)$ correction.

\noindent{\bf 2.} Still focussing on the $\epsilon = 0$ case, we shall consider the leading correction to the effective action that arises when the boundary condition $\lim_{z\to -\infty} \pi = \pi_*$ is relaxed. We shall consider $\lim_{z\to -\infty} \pi = \pi_*+\Delta \pi_\infty$ and treating $\Delta\pi_\infty$ as a small quantity we shall compute the correction to the dilaton
potential at leading linear order in $\Delta \pi_\infty$. Notice that with this modified boundary condition eqs.~(\ref{eq:ASolutionZero})-(\ref{eq:PhiSolutionZero})  will no longer be a stationary point, but that does not matter provided that we derive the resulting effective action keeping all effects. Indicating by $\Phi_0$ the $\epsilon=0$ solution and by $\psi_0$ and $\psi_{0KK}$ respectively the radion and the most general massive KK fluctuation around it, we can expand the action 
\beq
S(\Phi_0+\psi_0+\psi_{0KK}, \pi_*+\Delta\pi_\infty)\label{totalaction}
\eeq
in $\Delta \pi_\infty$ and in KK modes. As in the previous case, keeping $\psi_0$ as our low energy field and integrating out the massive KK modes, we find that the effect of the latter integration only starts at order  $(\Delta \pi_\infty)^2$ and at order $\Delta \pi_\infty \times \partial^2$. The leading $O(\partial^2)$ and $O(\Delta\pi_\infty)$ action is then simply
\beq
S(\Phi_0+\psi_0, \pi_*)+\Delta\pi_\infty \partial_{\pi_*}S(\Phi_0+\psi_0, \pi_*)\label{deltapi}
\eeq
The low energy effective theory described by the above Lagrangian,  will not possess Poincar\'e invariant solutions for $\Delta\pi_\infty\not = 0$. But as long as $\Delta\pi_\infty$
is small, the resulting solutions with non trivial spacetime dependent dilaton profile will give a valid effective 4D description of the corresponding 5D exact solutions.
In section \ref{subsec:radion1}, we shall directly compute the action of eq.~{\ref{deltapi}), while in section \ref{subsec:radion2} we shall deduce it indirectly by matching 5D solutions with AdS4 symmetry to the corresponding solutions in the 4D effective dilaton theory.

\noindent{\bf 3.} According to the dual 4D picture discussed in Section \ref{sec:4Dpicture} the potential at order $O(\Delta\pi_\infty)$ corresponds to the term $(\lambda-\lambda_*) 
\kappa'(\lambda_*)\varphi^4$. According to the discussion in that Section, once $\kappa'(\lambda_*)$ is known, RG considerations are sufficient to compute the dilaton mass at $O(\epsilon)$. This is the route we shall follow here, keeping in mind that in the 5D language RG invariance corresponds to the global dilatation diffeomorphism. We should also keep in mind that we can view this third step as the  addition of the $O(\epsilon)$ perturbation in eq. (\ref{totalaction}). As we did before one may worry about the effects
arising from integrating out the massive KK's. Again, by taking into account that the massive KK do not linearly mix with the dilaton in the unperturbed $\epsilon = \Delta\pi_\infty=0$ case, we conclude that these effects give terms that are at most $O(\epsilon \Delta\pi_\infty)$ and $O(\epsilon \partial^2)$. They therefore affect the radion squared mass only at order $\epsilon^2$. On the order hand, according to the discussion in Section \ref{sec:4Dpicture}, the $O(\Delta\pi_\infty)$ correction  in eq.~(\ref{deltapi}) gives a $m_\varphi =O(\epsilon)$.

\subsection{The dilaton mode at $\epsilon=0$} \label{subsec:radion}

Let us start from the case $\epsilon=0$ in which, as already said, we should find a vanishing radion-dilaton potential. This  is the step {\bf 1} we outlined above. A convenient  parametrization of the radion mode is given by the metric
\bea
ds^2&=&e^{2 \hat{A}(x,z)}\eta_{\mu\nu}dx^\mu dx^\nu-\hat{B}(x,z)^2 dz^2 \label{metricdilaton}\\
\hat{A}(x,z)&=&A_0(z+c(z)r(x))-r(x)/L, \label{dilaton0}\\ 
\hat{B}(x,z)&=&1+ c'(z) r(x),  \label{dilaton1}
\eea
and scalar field 
\beq
\hat\pi(x,z)=\pi_0(z+c(z)r(x))  \label{dilaton2}
\eeq
where $A_0$ and $\pi_0$ are the solutions in eqs.~(\ref{eq:ASolutionZero})-(\ref{eq:PhiSolutionZero}) for the choice  $z_*=0$, and $c(z)$ is a function such that $c(z_{IR})=0$ and $c(-\infty)=-1$.
Notice that, given the behaviour of  $\hat{A_0}(x,z)$ and $\hat\pi_0(x,z)$ at $z\to \infty$, the above mode has a finite action, i.e. it is normalizable. Moreover when $r$ is constant over spacetime the mode can be eliminated by the change of coordinates $\tilde z = z+c(z) r$, $\tilde x^\mu = e^{-r/L} x_\mu$, which does not affect the coordinate of the IR boundary and the asymptotic behaviour of the fields. We conclude that $r(x)$ has vanishing potential, and as such is a good interpolating field for the massless radion.
Notice that there remains some degree of arbitrariness in the choice of the radion wavefunction. 
 All functions $c(z)$ satisfying the same  boundary conditions should be equally good. Different choices of $c(z)$ will affect the mixing between massive KK's and the radion, and will be reflected in the $O(\partial^4)$ effective action, which we do not care about\footnote{There should however  exist a specific choice of $c(z)$ shuch that the quadratic mixing with the KK's vanishes \cite{Charmousis:1999rg}
.}. On the other hand we expect the leading $O(\partial^2)$ action to be unaffected by the freedom in the choice of $c$, as we shall now verify\footnote{In  Appendix \ref{app:gaugefixing} we shall discuss in more detail the realization of 4D dilations in the presence of a spacetime dependent $r$ and of KK excitations as well.}.

To compute the radion effective action, we simply plug eqs.~(\ref{metricdilaton})-(\ref{dilaton2}) into the action (\ref{eq:fullaction0}). The resulting expression is
\bea 
\frac{S}{M_5^3} &=& \int d^4 x \int_{-\infty}^{z_{IR}} dz \, \left\{ e^{2\hat{A}} \left[ -\frac{3}{2}\eta_{\mu\nu}(\p_\mu \hat{A})(\p_\mu \hat{B}) -\frac{3}{2}\eta_{\mu\nu}\hat{B}(\p_\mu \hat{A})(\p_\mu \hat{A}) + \frac{1}{2}\hat{B}\eta_{\mu\nu}(\p_\mu \hat{\pi})(\p_\nu \hat{\pi}) \right] \right.
\nonumber \\
&& \left.  + e^{4\hat{A}}\left[ 2\frac{  {\hat{A}'} {\hat{B}'}}{\hat{B}^2} - 5\frac{{(\hat{A}')}^2}{\hat{B}} - 2 \frac{ {\hat{A}''}}{\hat{B}} - \frac{ {(\hat{\pi}')}^2}{2\hat{B}} - \hat{B}V(\hat{\pi}) \right]   \right\} -\int d^4 x \, e^{4\hat{A}} \left. \left[ \frac{\tau_{IR}(\hat{\pi})}{2}-2 {\hat{A}'} \right]\right|_{z=z_{IR}}  \, .
\label{eq:radionAction}
\eea

By making use of the explicit expressions (\ref{dilaton0})-(\ref{dilaton2}) one finds the kinetic term:
\bea 
S_{kinetic} &=&(M_5 L)^3 \int d^4 x \frac{(\p_\mu r)^2}{2L^4} e^{-2(z_{IR}+r(x))/L}\, \times\,  Z(z_{IR} ) \label{eq:kineticterm} \\
Z(z_{IR} ) &=& \left( \frac{3}{2}+ 3e^{2z_{IR}/L} \int_{-\infty}^{z_{IR}} \frac{dz}{L} (e^{-2 z/L} - e^{2A_0(z)}) \right) \, ,
\nonumber
\eea
which does not depend on the form of $c$ as expected (we recall that $z_*$ has been set to zero without loss of generality). 
On the other hand by looking at the non-derivative interactions we can derive the radion potential, which can be written as a boundary term:
\be \label{eq:vanishingpotential}
S_{IR} = -\frac{(M_5 L)^3}{2}\int \frac{d^4 x}{L^4}\,  e^{4(A_0(z_{IR})-r/L)} \left( \tau_{IR}(\pi_0(z_{IR})) - 3 A_0'(z_{IR}) \right) \, .
\ee
This has precisely the expected form of a quartic term, as in (\ref{eq:withKappa}). The coefficient is however  exactly zero  thanks to the matching condition (\ref{eq:Matching1}).

The coefficient $Z$ for the kinetic term corresponds to the result in the RS model $Z=\frac{3}{2}$ up to a correction that measures the effect of the backreaction of $\pi$ on the metric.  Notice that $Z$ is always positive. It is natural to identify the dilaton with
\beq
\varphi =\frac{\sqrt{Z}}{L} e^{-(z_{IR}+r)/L}\label{dilatoncanonical}
\eeq
and to put into evidence the ``large N factor" $N^2\equiv (M_5 L)^3$ in the kinetic term
\beq 
{\cal L}_{kin}=\frac{N^2}{2} \partial_\mu \varphi\partial^\mu \varphi
\eeq
Moreover, under a dilation diffeomorphism $\tilde z = z+c(z) z_1$, $\tilde x^\mu = e^{-z_1/L} x_\mu$ the field $\varphi(x)$ does indeed transform as expected:
\beq
\varphi(x)\to \tilde \varphi(x)= e^{z_1/L} \varphi(e^{z_1/L} x)\, .\label{dilatonscale}
\eeq

It is interesting to ask how things change when $\epsilon\not = 0$. Naively it seems that by replacing $A_0$ and $\pi_0$ with the $\epsilon\not =0$ solution $(A,\pi)$   in eqs.~(\ref{metricdilaton})-(\ref{dilaton2}),  we can construct a mode that reduces to a change of coordinates at zero momentum. The problem with such a mode is that it is  not normalizable. The reason for that is the slow approach to the asymptote, now $\pi =0$, at the conformal boundary.
At $\epsilon\not = 0$  a  there isn't any normalizable mode  behaving like a pure change of coordinates at zero momentum, and thus we conclude that all modes are expected to have a potential. Pure changes of coordinate, however, still constrain the form of the resulting potential. Indicating by $A=A_0+\Delta A$ the warp factor in the $\epsilon \not = 0$ case, a normalizable mode interpolating for the dilaton  could now be written as in eq.~(\ref{metricdilaton}) with
\bea
\hat{A}(x,z)&=&A_0(z+c(z)r(x))+\Delta A(z+b(z) r(x))-r(x)/L, \label{dilaton0b}\\ 
\hat{B}(x,z)&=&1+ c'(z) r(x)\\  \label{dilaton1b}
\hat\pi(x,z)&=&\pi(z+b(z)r(x))  \label{dilaton2b}
\eea
where $c$ satifies the same boundary conditions as before, while $b$ coincides with $c$  at finite $z$, in particular $b(0)=0$,  but goes to  zero at $z\to -\infty$ fast enough to ensure normalizability. Since $b\not = c$ the diffeomorphism
 $\tilde z = z+c(z) z_1$, $\tilde x^\mu = e^{-z_1/L} x_\mu$ now changes the functional form of the asymptotic behaviour of the terms associated with $\Delta A$ and $\pi$. At lowest order in $z_1$ and $r$ we have
\beq
 \pi(z+ b(z) r)\to \pi(z-(c(z)-b(z))z_1 + b(z) (r-z_1))
\eeq
and similarly for $\Delta A$. Notice that in the asymptotic region $\Delta A$ can be expanded in a power series in $\pi$. Therefore asymptotically the above equation amounts to changing 
\beq
\pi(z) \to \pi (z+z_1)\label{piRG}
\eeq
We conclude that in the $\epsilon \not = 0$ case, eq.~(\ref{dilatonscale}) must be supplemented with the spurious transformation \ref{piRG} to leave the action invariant, with the obvious dual RG interpretation. In particular, in order  to respect the spurious scale invariance the potential must have the form 
\beq
\kappa (\pi(r))e^{-4r/L}\label{colemanweinberg}
\eeq
where $\pi(r)$ is  invariant under the combined action of \ref{piRG} and \ref{dilatonscale} (that is $r\to r-z_1$).

\subsection{Dilaton quartic: first approach} \label{subsec:radion1}

We now  carry out step {\bf 2} outlined in Section \ref{sec:5DsolutionEpdZero}. Still working at $\epsilon=0$ we compute the dilaton quartic at lowest order in the detuning parameter $\Delta \pi_\infty$. In order to do that we simply have to compute the dilaton action over a shifted $\pi$ background: in practice this amounts to taking
$\hat \pi = \pi_0(z+c r) + \Delta \pi_\infty$ in eq.~(\ref{dilaton2b}). Notice that such a shift has no effect in the bulk, as the action there only depends on $\partial \pi$. In particular the shifted fields are still a solution of the bulk equations of motion. The only contribution comes from the boundary tension, which at linear order in $\Delta \pi_\infty$ gives a dilaton potential
\beq
V= N^2 \Delta \pi_\infty \frac{\partial\tau_{IR}}{\partial \pi}\Big |_{\pi = \pi_{IR}}\frac{e^{4 (A(z_{IR})+z_{IR}/L)}}{2Z^2} \, \varphi^4
\eeq 
In terms of a general dilaton potential of the form $V = N^2 \kappa(\pi_\infty) \varphi^4$, this corresponds to 
\beq 
\frac{\partial \kappa}{\partial \pi_\infty}\Big |_{\pi_\infty = \pi_*} =  \frac{\partial\tau_{IR}}{\partial \pi}\Big |_{\pi = \pi_{IR}}\frac{e^{4 (A(z_{IR})+z_{IR}/L)}}{2Z^2} \, .\label{linearquartic}
\eeq
Using the expected general form \ref{colemanweinberg}  for the potential in the presence of a slowly evolving $\pi$, and carrying through precisely the same
reasoning that lead to eq.~(\ref{eq:mphiGenericDiscussion}) we find the dilaton squared mass at  leading $O(\epsilon)$:
\beq \label{eq:final:approach2}
m_\varphi^2 = \epsilon \,  P'(\pi_*) \, \tau'_{IR}(\pi(z_{IR}))\, \frac{e^{4 A(z_{IR})+2z_{IR}/L}}{2\, L^2\, Z(z_{IR}) }
\eeq
where we used also eq.~(\ref{betadual}) and eq.~(\ref{dilatoncanonical}).

\subsection{Dilaton quartic: second approach} \label{subsec:radion2}

From the discussion in Sections \ref{subsec:radion} and \ref{subsec:radion1}, interpreted from a purely 4D point of view, we deduce the dilaton Lagrangian:
\bea
\mathcal{L} &=& N^2 \left( 
\frac{1}{2} \, \partial_\mu \varphi\partial^\mu \varphi
- \kappa \, \varphi^4
 \right) \, . \\
\kappa &=& \Delta \pi_\infty \frac{\partial\tau}{\partial \pi}\Big |_{\pi = \pi_{IR}}\frac{e^{4 (A(z_{IR})+z_{IR}/L)}}{2Z^2} \label{eq:kappaSecondApproach} 
\eea
Following \cite{fubini}, this corresponds to a dilaton VEV with dS4 or AdS4 symmetry, of the type (\ref{pattern}).
It should then be possible to deduce the quartic coupling by solving the EOM in AdS5 with detuned asymptotic condition $\pi_\infty = \pi_* + \Delta\pi_\infty$, and then look at the curvature of the 4D sections.

We present in this Section this alternative approach.
We start from the metric:
\begin{eqnarray} \label{eq:metricWithCC}
ds^2 &=& e^{2 A(z)} g_{\mu\nu}(\bar\Lambda) dx^\mu dx^\nu - dz^2 
\end{eqnarray}
where:
\begin{equation} \label{eq:dSorAdS}
g_{\mu\nu}(\bar\Lambda) dx^\mu dx^\nu = 
\left\{
\begin{aligned}
& \frac{1}{(\sqrt{\bar\Lambda} t)^2} \eta_{\mu\nu} dx^\mu dx^\nu & \text{ (dS4) if } \bar\Lambda>0 \\
& \frac{1}{(\sqrt{-\bar\Lambda} x_3)^2} \eta_{\mu\nu} dx^\mu dx^\nu & \text{ (AdS4)  if } \bar\Lambda<0
\end{aligned}
\right. \, ,
\end{equation}
while the bulk field is $\pi = \pi(z)$.
The EOM now read \cite{Kaloper:1999sm}\cite{DeWolfe:1999cp}:
\begin{eqnarray}
\pi''+4 A' \pi' - \frac{\p V}{\p \pi} &=& 0 \label{eq:EOM1L}\\
 A'' + \bar\Lambda e^{-2A}+ \frac{2}{3}(\pi')^2&=&0 \label{eq:EOM2L} \\
(A')^2 -\bar\Lambda e^{-2A}+ \frac{1}{3}V(\pi) -\frac{1}{6}(\pi')^2&=&0 \, , \label{eq:EOM3L}
\end{eqnarray}
with the same matching conditions on the brane (\ref{eq:Matching1})-(\ref{eq:Matching2}).

To connect $\bar\Lambda$ with the quartic coupling $\kappa$ (\ref{eq:withKappa}) of the dilaton potential, we compare the curvature of the 4d sections. To be specific, if one starts from the dilaton field $\varphi$ in the 4D theory, with potential given by (\ref{eq:withKappa}), then  the metric that is seen by matter is:
\be
ds^2 = \frac{L^2 \, \varphi^2(x)}{Z}\, e^{2(A(z_{IR}) + z_{IR}/L)} \, \eta_{\mu\nu} \, dx^\mu \, dx^\nu
\ee 
where $\varphi(x)$ is given by (\ref{pattern}). 
Up to a change of coordinates, this is equivalent to (\ref{eq:dSorAdS}) with the identification:
\be \label{eq:identificationQuarticLambda}
\bar\Lambda = -\frac{2\kappa \, Z \, e^{-2z_{IR}}}{L^2} \, .
\ee

We now want to derive an expression for $\bar\Lambda $ by solving the EOM with detuned asymptotic value for the bulk scalar $\pi_\infty = \pi_* + \Delta\pi_\infty$, and then check that we recover (\ref{eq:kappaSecondApproach}) through (\ref{eq:identificationQuarticLambda}).
This computation is detailed in the Appendix \ref{appendix}.
Since we are interested in the solution close to the minimum of the radion potential, it is enough to solve the EOM at linear order in $\bar \Lambda$.
In principle however, to fully compute the potential, what one has to do is to find $\bar\Lambda$ by solving the EOM (\ref{eq:EOM1L})-(\ref{eq:EOM3L}).
In our case we define:
\bea
A(z) &\equiv& A_0(z)+\bar \Lambda \, \bar A_1(z) \\
\pi(z)= &\equiv& \pi_0(z)+\bar \Lambda \, \bar \pi_1(z)
\eea
and analogously for the derivatives.
We then impose the matching conditions (\ref{eq:Matching1})-(\ref{eq:Matching2}), that uniquely fix the values $\pi_*$ and $z_{IR}$ in the case $\bar \Lambda=0$.
Changing the asymptotic value of the field profile from $\pi_*$ to $\pi_\infty = \pi_* + \Delta\pi$ will require a non vanishing $\bar \Lambda$ in order for the IR matching conditions to be satisfied again, together with a shift $\Delta z$ in the position $z_{IR}$ of the IR brane.
At linear order in $\bar \Lambda$ one finds:
\bea
\delta z \,  \pi''_0+\bar \Lambda\, \bar \pi'_1&=&-\frac{1}{2}\frac{\p^2 \tau_{IR}}{\p \pi^2}\, \tilde\Delta\pi \nonumber \\
\delta z\, A''_0+\bar \Lambda\, \bar  A'_1&=&-\frac{2}{3}\pi'_0\, \tilde\Delta\pi \, , \label{matchingIR}
\eea
where $\tilde\Delta\pi=\Delta\pi+\Delta z\, \pi'_0+\bar \Lambda \bar \pi_1$.
Solving the system (\ref{matchingIR}) for the two unknowns $\bar \Lambda$ and $\Delta z $ we find:
\be \label{eq:resultLambdabar0}
 {L^2} \frac{ \bar\Lambda}{\Delta \pi_\infty} =
\frac{-2 \frac{\p\tau_{IR}}{\p\pi}\pi_0'' -3\frac{\p^2 \tau_{IR}}{\p \pi^2} A_0''}
{\bar  \pi_1 (2 \frac{\p\tau_{IR}}{\p\pi}\pi_0'' +3\frac{\p^2 \tau_{IR}}{\p \pi^2} A_0'') 
+\bar  \pi_1' (6 A_0'' -2 \pi_0' \frac{\p\tau_{IR}}{\p\pi})
+ \bar A'_1 (-6 \pi''_0 -3 \pi'_0 \frac{\p^2\tau_{IR}}{\p\pi^2})}
\ee
where all the functions are evaluated on the IR brane.
Using the zeroth order equations of motion (\ref{eq:Matching1})-(\ref{eq:Matching2}),
  (\ref{eq:resultLambdabar0}) simplifies as
\be \label{eq:finalFormula}
\frac{\bar\Lambda}{\Delta \pi_\infty} =
-\frac{1}{L^2}\times \left.
\frac{\frac{\p\tau_{IR}}{\p\pi}}{\frac{\p\tau_{IR}}{\p\pi} \bar \pi_1 - 3\bar  A_1'} \right|_{z_{IR}} \, .
\ee
%
Using the results in  Appendix \ref{appendix}, in particular (\ref{intformula}), one can straightforwardly show that this result is consistent with 
eq.~(\ref{eq:kappaSecondApproach}), through eq.~(\ref{eq:identificationQuarticLambda}).

%

\subsection{The limit of negligible backreaction} \label{subsec:radion0}

As a final check of our results, we consider the limit  of negligible backreaction where the computation is much simpler.
Working with $z_*=0$, this limit is obtained when $\tau_{IR}$ is such that the brane stabilizes at a position where:
\beq
\left| \frac{z_{IR}}{L}\right| \gg 1
\eeq
so that  
the local geometry is well approximated by AdS5 everywhere, that is $A(z)\simeq  -z/L$.

More precisely,  
starting from a zeroth order solution with $\pi=\pi_{IR}$ and $A(z)= -z/L$, corresponding to $f=0$, $f'=0$ and $\epsilon=0$, we
we can solve the EOM in an order by order expansion in the latter three quantities, treted as small parameters.
Indeed by considering the correction $\Delta T$ to the energy momentum tensor that come from $\pi$ and from the IR brane tension, and demanding that it be subdominant to the bulk cosmological constant, one indeed obtains
\beq
 f(\pi_{IR})\ll 1 \qquad ( f^\prime(\pi_{IR}) )^2\ll1 \qquad \epsilon \ll 1
\eeq
where the second condition comes from the $(\pi^\prime)^2$ contribution to $\Delta T$ after having taken the IR brane matching condition into account. Notice that, strictly speaking, provided a point where both $f$ and $f^\prime$ vanish exists, eqs.~(\ref{eq:EOM1})-(\ref{eq:Matching2}) are endowed with the unperturbed slice of AdS solution. In this respect we do not need to impose that higher derivatives of $f$ be small. In particular $f''$ could be $O(1)$.
Obviously the existence of a value of $\pi$ such that  $f=f^\prime=0$ requires tuning: a tuning that 
only lends us some computational hedge, but which is conceptually not needed.

%
%

Let us focus for definiteness on the case of a quadratic bulk potential $P(\pi)=-2\pi^2$. Once $\epsilon \not = 0$ the leading order $\pi$ solution is given over all space (see eq.~(\ref{eq:AsymptPhiSolution}))
\begin{equation} \label{GWpi}
\pi = \pi_* e^{\epsilon z/L}\, + \hat{\pi}_{*} e^{(4-\epsilon)z/L}\, .
\end{equation}
In order to make contact with eq.~(\ref{dilatoncanonical}) we parametrize the radion  by considering a displaced brane at the position $z_{IR}+r$. The matching condition (\ref{eq:Matching2}), then reads 
\be \label{eq:matching:noback}
\pi(z_{IR}+r) = -\frac{1}{2(4-\epsilon)} \frac{\p f}{\p \pi} (\pi(z_{IR}+r)) + \frac{4-2\epsilon}{4-\epsilon} \, \pi_* \, e^{\epsilon (z_{IR}+r)/L} \, \equiv g(\pi_* e^{\epsilon (z_{IR}+r)/L}) \, .
\ee
Notice that in the limit $\epsilon =0$, $f(\pi_{IR})=0$ this condition implies $\pi(z)=\pi_{IR}=const$.

Indicating the canonical dilaton 
as in eq.~(\ref{dilatoncanonical}), with now $Z\simeq \frac{3}{2}$, and substituting the solution of the EOM into the action one obtains:
\beq \label{eq:effActionSmallBackreac}
S = N^2 \, \int d^4 x \, \left(  \frac{1}{2}(\partial \varphi)^2   -\varphi^4 \kappa(\pi(\varphi)) \right)
\eeq
where, keeping only the terms that are not suppressed by powers of $\epsilon$:
\be \label{eq:dilatonpotential:noback}
\kappa(\pi(\varphi)) =  \frac{1}{2 Z^2}\left( f(g(\pi(\varphi))) -\frac{1}{2}\left( g(\pi(\varphi)) - \pi(\varphi) \right) \, \frac{\p f}{\p \pi}(g(\pi(\varphi))) \right) \, ,
\ee
and we defined the running coupling:
\be
\pi(\varphi) = \pi_* \left( \frac{L}{ Z^{\frac{1}{2}} } \varphi \right)^{-\epsilon} \, .
\ee
Notice that the term $\pi(\varphi)$  subtracted from $g(\pi(\varphi))$ in the second term arises from the integration in the region $z\to -\infty$. The reason for this contribution is that in the computation a la GW we did not parametrize the radion with localized mode: the profile of $\pi$ towards the conformal boundary depends on $r$.
Now,  since:
\be
g(\pi(\varphi)) - \pi(\varphi) = -\frac{\epsilon}{4}\pi(\varphi)-\frac{1}{2(4-\epsilon)} \frac{\p f}{\p \pi} (g(\pi(\varphi)))
= -\frac{1}{8}  \frac{\p f}{\p \pi} (g(\pi(\varphi)))+ O(\epsilon)
\ee
we conclude that, at leading order in $\epsilon$ the above result coincides with:
\be
\kappa(\pi(\varphi)) =\frac{1}{2Z^2}\left (f(g(\pi(\varphi)))+\frac{1}{16}( \, f'(g(\pi(\varphi))) \, )^2\right )
\ee
We can quickly compare this result  with eq.~(\ref{linearquartic}) by considering the limit $\epsilon =0$ and taking the first derivative with respect to $\pi_*$. In agreement with eq.~(\ref{linearquartic}) we find:
\beq
\frac{\partial \kappa}{\partial \pi_*}=\frac{1}{2Z^2}\left (f'(g(\pi_*))+\frac{1}{8}f''(g(\pi_*))f'(g(\pi_*))\right )\frac{\partial g}{\partial \pi_*}=\frac{1}{2Z^2}f'(g(\pi_*))\equiv \frac{2}{9}f'(\pi_{IR})\label{quarticnobac}
\eeq
where we used eq.~(\ref{eq:matching:noback}) to derive:
\beq
\frac{\partial g}{\partial \pi_*}=\frac{1}{1+\frac{1}{8}f''(g(\pi_*))}\, .
\eeq
The consistency with our previous results for the mass of the dilaton at leading order in $\epsilon$ follows straightforwardly. We should notice that by considering eqs.~(\ref{eq:EOM3})-(\ref{eq:Matching2}),
in the limit $\epsilon =0$ one obtains the following condition on $\pi_{IR}$:
\beq
f(\pi_{IR})+\frac{1}{16}(f'(\pi_{IR}))^2-\frac{1}{6}(f(\pi_{IR}))^2=0
\eeq
that would coincide with  eq.~(\ref{quarticnobac}) if it wasn't for the $f^2$ term. This is consistent with our leading approximation, where $f$ and $(f')^2$ are independent corrections to the energy momentum tensor, and should be considered as equally important. However $(f)^2$ is subdominant to $f$. In order to consistently take those higher orders into account we would have  to consider the backreaction of the metric.
\section{Conclusions} \label{sec:conclusions}

The issue of spontaneus breaking of the conformal group $O(N,2)$ in N-dimensional quantum field theory resembles very closely the cosmological constant problem in general relativity.
In the former case the symmetry of the system does not forbid the associated Goldstone boson (dilaton)  to have a quartic potential. The resulting pattern of symmetry breaking to either deSitter or  anti-deSitter or Poincar\'e subgroups of $O(N,2)$ then follows respectively from the choice $\kappa<0$, $\kappa>0$ and $\kappa=0$ for the quartic potential. In particular the Poincar\'e subgroup is selected only in a subset of measure zero of the parameter space. It thus appears rather non generic.
In the case of gravity, the similar pattern emerges because the relevant symmetry, diffeormophism invariance, does not prevent the presence of a potential term, the cosmological constant $\Lambda$, for the associated gauge field. The (maximally symmetric) solutions are then either deSitter, anti-deSitter or Poincar\'e, depending on $\Lambda>0$, $\Lambda<0$ or $\Lambda =0$. The cosmological constant problem lies in the apparent non genericity of the choice $\Lambda=0$.
As nicely elucidated by Sundrum \cite{Sundrum:2003yt}, the analogy between the two problems should not come out as surprising. Indeed the non linear realization of $O(N,2)$ through a dilaton, represents a possible relativistic extension
of Newtonian gravity, the so-called theory of scalar gravity.

In this paper we presented a scenario, based on effective  field theory, that produces a pseudo-dilaton with a naturally small potential. The two key features  to achieve that are
\begin{itemize}
\item The existence of a ``landscape'' of values for the quartic coupling $\kappa$ of the effective dilaton potential, containing the point $\kappa=0$.
\item The explicit breakdown of conformal invariance by a naturally small parameter, associated with a nearly marginal coupling
\end{itemize}
The combination of the above two key features  gives rise to a specific vacuum dynamics  according to which the minimum robustly sits near the point $\kappa=0$.
We discussed in detail a 5D holographic realization that explicitly illustrates our solution is {\it natural}  according to the standard naturalness criterion \cite{'tHooft:1979bh}. The smallness  of the dilaton potential around its minimum directly follows from the presence in the 5D bulk of the  pseudo-Goldstone boson of an internal symmetry. Our model represents a variant of the Goldberger-Wise (GW) mechanism \cite{Goldberger:1999uk} of radion stabilization in the Randall-Sundrum model \cite{Randall:1999ee}. The novelty, with respect to previous implementations of the GW mechanism, is that our construction makes clear that in order to obtain a light radion there is no need to tune, even approximately, the tension of the IR brane. As long as the bulk potential for the GW scalar is flat, the minimum of the radion potential arises at a point where the overall potential is small, as if the IR brane tension were practically tuned. Aside this result we checked that our model is sensible in that there are no ghosts or tachyons.

It is now clearly interesting to ask what our example could teach us on the true cosmological constant problem,
the one concerning (quantum) gravity. Even before trying to think of an explicit 4D analogue, one interesting step would be to understand how our mechanism can be embedded in a thermal history of a scalar gravity toy universe. Some basic questions would be: Under what initial conditions is the dilaton driven at late times to the region where its potential is small? What would such a cosmology look like? Would there be the analogue epochs of radiation domination, matter domination and  structure formation then followed by a period of accelerated expansion? In order for our example to be interesting, one would like to exclude the need for extreme tuning of initial conditions in order to achieve such a cosmology. We understand that Sundrum, in a forthcoming paper, has gone  a long way towards addressing these issues \cite{SundrumInPreparation}. Encouraged by that result, one may speculate about a real gravity translation of our scenario.
Even allowing for some loss in translation, what could  our two key features  look like? A landscape of values for the cosmological constant could for instance be provided by the a 3-form field $A$ and its 4-form field strength $F=dA$ (see for instance \cite{Weinberg:1988cp}). Indeed $F$ has a continuum of constant vacuum solutions. The effective cosmological constant is just a function $\Lambda(F)$ and we generically expect $F_*$ esists such that $\Lambda(F_*)=0$. Notice, in passing, that $F\not = 0$ provides the {\it smallest} ``higgsing'' of gravity, where the group of diffeomorphisms is spontaneously broken to the subgroup of volume preserving diffeomorphisms. The freedom in the choice of $F$, which can be associated with the choice of boundary conditions, directly corresponds to the freedom in the choice of $\Lambda$ in unimodular gravity. In view of these analogies, the field strength $F$  encouragingly looks 
like the real gravity analogue of the marginal coupling $\lambda$ of our toy gravity construction. How could we then mimick the second key  ingredient? Like in the toy example, we need to somehow  lift  the degeneracy over the landscape. One possibility to achieve that  was indeed pointed out some time ago by Brown and Teitelboim (BT) in a visionary paper \cite{Brown:1988kg}. BT have shown that when there exist 2-branes that couple to $A$,  brane nucleation by quantum tunnelling can relax by discrete jumps the value of $F$. If the brane tension is small enough the succesive jumps could then relax the effective cosmological constant down to its observed value. However, in the range of parameters where this can happen, the rate of bubble nucleation is so small that it seems difficult to implement the mechanism in a realistic cosmology (see however ref. \cite{Feng:2000if} for a broader perspective). Perhaps, leaving quantum tunnelling aside, another perspective to eliminate the degeneracy of the $F=const$ solutions and realize a surrogate of our second key feature, would be to ``higgs" the gauge symmetry associated with the 3-form and give it a mass. That is done by adding a 2-form $B$ that shifts under the gauge transformation $B\to B+\alpha$ whereas $A$ transforms like $A\to A+d\alpha$. 
The presence of a mass term would give rise to a slow evolution of the field strength $F$ that could, in principle, relax towards the value $F_*$ where the cosmological constant vanishes. In a sense the addition of a Goldstone 2-form $B$ classically screens the field strength $F$, where  BT brane nucleation  did so by quantum tunnelling. The analogy 
with the quantum screening mechanism suggests that  a massive 3-form could be a promising direction. On the other hand, from another perspective, a massive 3-form is dualized as a massive scalar, so that in the limit of small mass the system should just corresponds to a version of quintessence (see for instance \cite{Koivisto:2009fb}). From the latter perspective it would seem tuning is still needed to achieve a cosmology with small effective cosmological constant at late times, but perhaps  an explicit investigation is warranted. 

To conclude: a 4-form field strength could indeed play the role  of our marginal coupling, however we have yet to identify a successful analogue of our second key feature. In scalar gravity that role was played by an explicit small breaking of the relevant global symmetry, conformal invariance. In the case of real gravity the relevant symmetry is a gauged one, for which there is no analogue of explicit breaking: breaking diffeomorphism invariance invariably brings in new degrees of freedom, thus entering the mine field of {\it modified gravity}. Is there a way out?
%

\section*{Acknowledgments}

RR thanks Roberto Contino and Alex Pomarol for developing the original idea\cite{cpr} that lead to this paper.
DP and RR also thank Raman Sundrum for engaging discussions on the scalar guise of the cosmological constant problem.
This research is supported by the Swiss National Science Foundation under contract 200021-125237.
The work of D.P. is supported by the NSF Grant PHY-0855653.

\appendix

\section{KK modes and scale invariance}  \label{app:gaugefixing}

In this Appendix we discuss in more detail the vanishing of the dilaton potential and dilatations in the presence of KK excitations.
The most general metric compatible with the gauge choice $g_{\mu 5}=0$ can be conveniently written as 
\be \label{metricdilatonApp}
ds^2 = e^{-2 r(x)/L } \left[ e^{2 A(z+c(z)r(x))}\eta_{\mu\nu} + h_{\mu\nu}(x,z) \right] dx^\mu dx^\nu-\left[  1+ c'(z) r(x)  \right]^2 e^{2\phi(x,z)}dz^2 
\ee
while the scalar field is:
\be
\hat\pi(x,z)=\pi_0(z+c(z)r(x)) + \chi(x,z)\label{piApp}
\ee
Notice that the condition $g_{\mu 5}=0$ does not completely fix the gauge, one more condition being needed, for instance $\chi = 0$.
Now, for $\Psi_{KK}\equiv (h_{\mu\nu},\phi,\chi)=0$ this field configuration reduces to the dilaton of eqs.~(\ref{metricdilaton})-(\ref{dilaton1}). Moeover for $r(x)={\rm const}$ the dependence on $r$ can clearly be completely  transferred to the KK modes  via a diffeormorphism $z+c(z)r =\bar z$, $e^{-r}x=\bar x$. This fact, together with the stationarity of the action around the solution, implies that any $r^n \Psi_{KK}$ mixing term vanishes at zero momentum. Indeed   because of  Lorentz invariance the mixings   must  be $O(\partial^2)$. We are thus reassured that by integrating out the KK modes at tree level we only affect the dilaton action at $O(\partial^4)$: the potential at $\epsilon =0$ vanishes. Notice however that 
terms of quadratic and higher order in $\Psi_{KK}$, in particular KK masses, will instead depend on the constant mode of $r$. This means that, in general, quantum corrections will affect the dilaton potential. That is not unexpected, but also not worrysome. For instance the bulk and brane tensions will be modified at the quantum level. Therefore, unless we  modify the asymptotic behaviour of $\pi$ accordingly, we shall not have a Poincar\'e invariant solution, corresponding to a non vanishing dilaton quartic. Reasoning with the 4D dual picture it is pretty evident that a suitable shift of $\pi_*$ that does the job must exist.

It is interesting to study in detail the transformation of the fields under dilatations, as partially done in eq.~(\ref{dilatonscale}). One is easily convinced that by defining
\beq
\tilde z = z+c(z) z_1\qquad\qquad \tilde x^\mu = e^{-z_1/L} x_\mu \label{scaleapp}
\eeq 
 the metric and scalar field in the new coordinates read
\bea 
ds^2 &=& e^{-2 \tilde r(\tilde x)/L } \left[ e^{2 A(\tilde z+\tilde c(\tilde z)\tilde r(\tilde x))}\eta_{\mu\nu} + h_{\mu\nu}(x,z) \right] d\tilde x^\mu d\tilde x^\nu-\left[  1+ \tilde c'(\tilde z) \tilde r(\tilde x)  \right]^2 e^{2\phi(x,z)}d\tilde z^2 \\
\hat\pi(x,z)&=&\pi_0(\tilde z+\tilde c(\tilde z)\tilde r(\tilde x)) + \chi(x,z)
\eea
where $\tilde r(\tilde x)\equiv r(x)-z_1$ and $\tilde c(\tilde z)\equiv c(z)$. Notice that we knowingly left the dependence of $h_{\mu\nu},\phi,\chi$ on $(\tilde x,\tilde z)$ to be implicit, that is via the dependence of $(x,z)$ on $(\tilde x,\tilde z)$. Notice also that $\tilde c(\tilde z)\not = c(\tilde z)$, so that the parametrization of the radion mode is different in the new coordinates.  However, from the definition $\tilde c(\tilde z)=c(z)$ and the first of eq.~(\ref{scaleapp}), one concludes $\tilde c$ satisfies the same boundary conditions $\tilde c(0)=0$, $\tilde c(-\infty)=-1$. It represents thus another, and equally good, parametrization of the dilaton. The change in the radion wavefunction can however be traded for a shift of the KK modes. By defining
\bea
\tilde h_{\mu\nu}(\tilde x, \tilde z)& \equiv& h_{\mu\nu}(x,z) +\left (e^{2A(\tilde z+\tilde c(\tilde z)\tilde r(\tilde x))-2A(\tilde z+ c(\tilde z)\tilde r(\tilde x))}-1\right )\eta_{\mu\nu}\\
\tilde \phi(\tilde x, \tilde z)& \equiv&\phi( x,  z)+\ln\frac{1+\tilde c'(\tilde z)\tilde r(\tilde x)}{1+ c'(\tilde z)\tilde r(\tilde x)}\\
\tilde \chi(\tilde x, \tilde z)& \equiv&\chi( x,  z)+\pi_0(\tilde z+\tilde c(\tilde z)\tilde r(\tilde x))-\pi_0(\tilde z+ c(\tilde z)\tilde r(\tilde x))
\eea
the metric and scalar field are written in terms of $\tilde r(\tilde x), \,\tilde h_{\mu\nu}(\tilde x, \tilde z), \,\tilde \phi(\tilde x, \tilde z),\, \tilde \chi(\tilde x, \tilde z)$ in the same form as 
eqs.~(\ref{metricdilatonApp},\ref{piApp}). At this stage to complete the discussion we should find a gauge fixing condition 
for  the remaining one 5D degree of freedom that is stable under  the above 
transformation law. This can easily be done. For rinstance, at the linearized level, a suitable gauge is given by 
\be
\frac{d}{dz} \left( \frac{h_\mu^\mu}{8 A'} \right) = \phi\, .
\ee

\section{Kaluza-Klein decomposition and sanity-check of the spectrum} \label{appendix:KK}

In this Appendix we briefly discuss under which conditions there are no tachyons in the spectrum for $\epsilon = 0$, and we show that there is only one massless mode. This is enough to prove that there are no tachyons also for a small scalar bulk field mass $\epsilon$, because all the correction to the masses are analytic in $\epsilon$, and the dilaton mass at $O(\epsilon)$ can generically be positive.
The absence of ghosts can be shown through an explicit diagonalization of the Lagrangian, as done for example in \cite{Kofman:2004tk} or \cite{Kiritsis:2006ua}.

It is convenient to work in conformally-flat coordinates. The background metric and field can be written as:
\begin{equation}
\begin{split}
& d s^2 = a^2(y) \left[ d x^\mu d x_\mu + d y^2 \right]
\\ & \varphi(x, y) = \varphi_0 (y) \, .
 \end{split}
\end{equation}
The most general set of linear perturbations is the following:
\begin{equation}
\begin{split}
& d s^2 = a^2(y) \left[ d x^\mu d x^\nu (\eta_{\mu \nu} + h_{\mu \nu}) + A^\mu dx_\mu d x^5 + (1 + 2 \phi)d y^2 \right]
\\ & \varphi(x, y) = \varphi_0 (y) + \chi(x,y)
 \end{split}
\end{equation}
with the following properties under gauge transformation $\delta x^\mu = \xi^\mu$, $\delta y = \xi^5$:
\begin{eqnarray}
 \delta h_{\mu \nu} &=& -\partial_{\mu} \xi_{\nu} -\partial_{\nu} \xi_{\mu} - 2 \eta_{\mu \nu} \frac{a'}{a} \xi^5 \nonumber\\ 
\delta A_\mu &=& - \xi_\mu' - \partial_\mu \xi^5 \\
\delta \phi &=&  - {\xi^5}' - \frac{a'}{a} \xi^5 \nonumber \\ 
\delta \chi &=& - \varphi_0' \xi^5 \, . \nonumber
\end{eqnarray}
In order to find the equations of motions it is convenient to choose the gauge:
\begin{eqnarray}
\partial_\mu A^\mu &=& 0
\\ \nonumber h_{\mu \nu} &=& h^{T T}_{\mu \nu} + 2 \eta_{\mu \nu} \psi
\end{eqnarray}
where $h^{T T}_{\mu \nu}$ is transverse-traceless and $\psi = \frac{1}{8} h^\mu_\mu$. 
The bulk Lagrangian for this perturbations is:
\begin{equation}
\begin{split}
\mathcal{L} = & \frac{1}{2} a^3 \left[ L^{(2)}_{ein} -\frac{1}{4} h'^{\rho \sigma} h'_{\rho \sigma} + 16 \psi'^2 - \frac{1}{4} F^{\mu \nu} F_{\mu \nu} + \partial_\mu \chi \partial^\mu \chi - \chi'^2 + 6 \partial_\mu \phi \partial^\mu \psi \right. \\ & \left.
 + 2 \varphi_0' \phi' \chi + 8 \varphi_0' \psi' \chi + 4 \varphi_0' \phi \chi' + 3 \frac{a'}{a} \left( -2 \phi \phi' - 8 \phi \psi' \right) \right]
\end{split}
\end{equation}
where $L^{(2)}_{ein}$ is the usual kinetic term for the graviton and $F_{\mu \nu} \equiv \partial_\mu A_\nu - \partial_\nu A_\mu$.

The EOM for the tensor and vector modes are:
\begin{eqnarray}
h_{\mu \nu}^{T T ''} + 3 \frac{a'}{a} h_{\mu \nu}^{T T '} + \square h_{\mu \nu}^{T T} &=& 0 \\
 \square A_\mu^T &=& 0 \\
  \left( a^3 A_{\mu}^T \right)' &=& 0
\end{eqnarray}
where $A_{\mu}^T$ is the transverse component of $A_{\mu}$. These equations imply that we have massive (non-tachyionic) gravitons and a massless graviton. The massless vector field $A_{\mu}^T$ is eliminated by the Dirichlet boundary condition on the brane.

In the scalar sector the perturbations $\psi$, $\chi$ and $\phi$ are not independent, due to the non-dynamical Einstein equations:
\begin{eqnarray}
\label{eq:constraints}
3 \psi' + \varphi_0' \chi -3 \frac{a'}{a} \phi &=& 0 \\
\phi &=& -2 \psi
\end{eqnarray}
The dynamical EOM for the perturbation $\psi$ can be written in the following form (see e.g. \cite{Lesgourgues:2003mi}):
\begin{eqnarray}
\label{eq:eqpsi1}
\Pi' &=& \left(- 7 \frac{a'}{a}\right) \Pi + \left(-m^2 + \frac{2}{3} \varphi_0'^2 \right) \psi \\
\label{eq:eqpsi2}
\psi' &=& \Pi - 2 \frac{a'}{a} \psi
\end{eqnarray}
The boundary condition on the infrared brane at $y=y_{I R}$  is:
\begin{eqnarray}
\label{eq:boundary}
0 &=& g_+ \Pi (y_{I R}) + m^2 \psi (y_{I R})   \\
g_+ &\equiv&  4 \frac{a'}{a}(y_{I R}) - \frac{1}{2} a(y_{I R}) \, \tau_{IR}''(\varphi_0(y_{I R})) \, .
\end{eqnarray}
where $\tau_{IR}$ is the tension on the infrared brane. 
 For the following discussion it is important to note that $\frac{a'}{a} < 0$ in the whole space and that the kinetic part of the Lagrangian for the scalar perturbations, after eliminating $\phi$ through the constraint equations, reads:
\begin{equation}
\mathcal{L}_{kin} = \frac{1}{2} \int d^4x \,  d y \, a^3 \, \left[ 6 \partial_\mu \psi \partial^\mu \psi + \partial_\mu \chi \partial^\mu \chi \right] \, .
\end{equation}
In order to have a physical scalar mode, the kinetic terms must be normalizable in the fifth dimension:
\begin{eqnarray}
\int_{0}^{y_{I R}} d y \, a^3 \psi^2 & < &  \infty
\\ \nonumber \int_{0}^{y_{I R}} d y \, a^3 \chi^2 & < &  \infty
\end{eqnarray}

\subsection{Massive spectrum}

We now slightly modify the argument given in \cite{Lesgourgues:2003mi} to find a sufficient condition for the positivity of the spectrum.
In the asymptotic region $y \sim 0$, we have $a \sim \frac{1}{y}$, and thus (\ref{eq:eqpsi1}) and (\ref{eq:eqpsi2}) reduce to:
\begin{equation}
\psi''-\frac{9}{y} \psi'+  \frac{16}{y^2} \psi + m^2 \psi = 0
\end{equation}
Let us suppose that $m^2 < 0$, and we want to find a contradiction. There are two independent solutions for this differential equations, with the following leading behaviors in the UV:
\begin{eqnarray}
\psi_1  \sim  y^2 + o(m^2 y^4) & , & \chi_1 \sim 1 \\ \nonumber
\psi_2  \sim  y^8  & , & \chi_2 \sim y^4
\end{eqnarray}
where $\chi_1$, $\chi_2$ are derived from (\ref{eq:constraints}). The first solution has a non-normalizable kinetic term for $\chi_1$, and thus it must be $\psi \sim \psi_2$. This function is such that $\psi$ and $\Pi$ (calculated from (\ref{eq:eqpsi2})) have the same sign in the UV. We can then assume $\psi > 0$, $\Pi > 0$ in the UV without loss of generality. 
Furthermore (\ref{eq:eqpsi1}) and (\ref{eq:eqpsi2}) imply that $\Pi ' > 0$ and $\psi' > 0$, upon plugging in the EOMs for $\epsilon = 0$, and so $\Pi$ and $\psi$ keep the same sign in the whole space. Thus (\ref{eq:boundary}) cannot hold on the IR brane if $g_+  < 0$. This implies that a sufficient condition for the positivity of the spectrum is:
\begin{equation}
\frac{1}{2} a(y_{I R}) \, \tau_{IR}''(\varphi_0(y_{I R})) > 4 \frac{a'}{a}(y_{I R})
\end{equation}
which is always true for $\tau_{IR}'' > 0$.

\subsection{Counting of massless modes}

We can now check that for $\epsilon = 0$ there is one massless mode, which we identify as the dilaton.
In fact for $m^2 = 0$ Eqs. (\ref{eq:eqpsi1}) and (\ref{eq:eqpsi2}) can be solved exactly:
\begin{equation}
\psi = C_1 \frac{a'}{a^4} + C_2 \left( 1 - 2 \frac{a'}{a^4} \int a^3 d y \right) \equiv C_1 \psi_1 + C_2 \psi_2
\end{equation}
where $C_1$ and $C_2$ are two integration constants.
The leading behaviors of these modes in the UV are:
\begin{eqnarray}
\psi_1  \sim  y^2 + o(y^{10}) &, & \chi_1 \sim y^6 \\ \nonumber
\psi_2  \sim  y^8 &,&  \chi_2 \sim y^4
\end{eqnarray}
both of which give normalizable kinetic terms. 
The IR boundary condition can thus be satisfied by a combination of the two solutions, and there is one massless mode in the spectrum.
This completes the proof.

\section{Solutions of the EOM at first order in $\Lambda$} \label{appendix}

We have to solve the EOM (\ref{eq:EOM1L})-(\ref{eq:EOM3L}) with $\epsilon=0$ and matching conditions (\ref{eq:Matching1})-(\ref{eq:Matching2}).
Eq. (\ref{eq:EOM1L}) is equivalent to:
\be
\pi'(z)=C e^{-4A(z)}
\ee
which substituted into (\ref{eq:EOM3L}) gives, at first order in $\bar\Lambda$:
\be\label{expansion}
-4 ~dz=\frac{dy}{\sqrt{1+\bar a y^{3/2}+y^2}}\approx\frac{dy}{\sqrt{1+y^2}}-\frac{\bar a }{2}\left(\frac{y}{1+y^2}\right)^{3/2} dy
\ee
with $y\equiv (\sqrt6/|C|) e^{4A}$ and $\bar a\equiv L^2 \, \bar\Lambda (\sqrt6/|C|)^{1/2}$. Notice that the expansion in (\ref{expansion}) works in the region $\bar a\lesssim1$. Here, the term $\bar a y^{3/2}$ is always subdominant, since in the region $y< \bar a^{-2/3}$ it is $1>\bar a y^{3/2}$, while in the region $y>\bar a^2$ it is $y^2>\bar a y^{3/2}$, and the two regions overlap.
Integrating both sides from 0 to $y$ one gets:
\be\label{eq1}
-4(z-z_*)/L 
\equiv f(y)+\bar a \,  g(y),
\ee
where $z_*$ is the same as in (\ref{eq:ASolutionZero}), and we choose our coordinates such that $z_*=0$, while:
\bea
f(y)&=&{\rm arcsinh} ~y\\
g(y)&=&\frac{1+2y^2}{2y^{3/2}\sqrt{1+y^2}}-\frac{\sqrt{1+y^2}}{2 y^{3/2}}{}_2F_1(-y^2;-\tfrac{1}{4},1,\tfrac{1}{4}).
\eea
In order to invert these relations and get $y$ as a function of $z$ we write, always at linear order in $\bar a$:
\bea
y(z) & = & F(z)+\bar aG(z) \label{eq2} \\
-4z/L &=& 
f(F(z))+\bar a f'(F(z))G(z)+\bar a g(F(z)).
\eea
From the $O(\bar a^0)$ we have $f(F(z))=-4z/L$, while at $O(\bar a)$ we find:
\be\label{eq3}
G(z)=-\frac{g(F(z))}{f'(F(z))}.
\ee
In conclusion:
\be\label{eq4}
F(z)=\sinh(-4z/L),\quad f'(F(z))=\frac{1}{\cosh(-4z/L)}
\ee
and:
\be
y(z)=\sinh(-4z/L)-\bar a g(\sinh(-4z/L))\cosh(-4z/L).
\ee
This implies that the solutions we are looking for are given by:
\bea
A(z)&=&\frac{1}{4}\left(\log\frac{|C|}{\sqrt6}+\log\sinh(-4z/L)-\bar a g(\sinh(-4z/L))\coth(-4z/L) \right)
\label{eq5} 
\eea
and:
\be\label{eq6}
\pi'(z)= {\rm sgn}C~\sqrt 6\left[\frac{1}{\sinh(-4z/L)}+\bar a g(\sinh(-4z/L))\frac{\cosh(-4z/L)}{\sinh^2(-4z/L)}\right]
\ee
where with the prime we still mean the derivative with respect to $z$.
Using:
\be
\int_\zeta^\infty\frac{dx}{\sinh x}=-\log\tanh\frac{\zeta}{2}
\ee
and:
\be
\int_\zeta^\infty~g(\sinh x)\frac{\cosh x}{\sinh^2 x}d x=\int_{\sinh \zeta}^\infty ~\frac{g(w)}{w^2}dw\equiv H(\sinh\zeta)
\ee
 one also finds:
\be\label{fieldexpansion}
\pi(z)=\pi_\infty+\frac{{\rm sgn}~C}{2}\sqrt{\frac{3}{2}}\left[-\log\tanh\frac{(-4z/L)}{2}+\bar a H(\sinh(-4z/L))\right]
\ee
Notice that $H(\sinh\zeta)\to 0$ as $\zeta\to \infty$. 

Finally, in order to connect $\bar a$ with $\bar \Lambda$ we need to specify the value of the constant $C$ in (\ref{eq5}). For our purposes, since we work at linear order in $\bar a$, it is enough to impose that at zeroth order in $\bar a$ we have $A(z)\sim -z/L$ for $|z|\gg L$. This fixes $|C|=2\sqrt{6}$ so that, at the end of the day, for our purposes we can take $\bar a = L^2\,  \bar\Lambda / \sqrt{2}$.

As a final comment, the fact that (\ref{eq:finalFormula}) agrees with (\ref{eq:kappaSecondApproach}) can be checked by using:
\begin{equation}
\label{intformula}
\int_{-\infty}^{z_{IR}} dz (e^{-2 z/L} - e^{2A_0(z)}) = -\frac{1}{2} e^{-2 z_{IR}/L} +\frac{\sqrt{2}}{2} \frac{t^{\frac{3}{2}}}{\sqrt{1+t^2}} -\sqrt{2} H(t) + \frac{\sqrt{2}}{t} g(t),
\end{equation}
where
\begin{equation}
t = \sinh (-4 z_{IR} /L)
\end{equation}
and the function $A_0(z)$ is defined by (\ref{eq:ASolutionZero}) with $z_*=0$.

\vspace{0.3cm}




\begin{thebibliography}{nn}
{\small

\bibitem{cpr}
The main idea discussed in this paper was originally developed by R. Contino, A. Pomarol and R. Rattazzi
 in unpublished work,
 see talk by R. Rattazzi at Planck 2010, CERN [\href{http://indico.cern.ch/getFile.py/access?contribId=163\&resId=0\&materialId=slides\&confId=75810}{indico/contribId=163\&confId=75810}]. See also talk by A. Pomarol at Xmas10 \href{http://www.ift.uam-csic.es/workshops/Xmas10/doc/pomarol.pdf}{http://www.ift.uam-csic.es/workshops/Xmas10/doc/pomarol.pdf}


\bibitem{Salam:1970qk} 
  A.~Salam and J.~A.~Strathdee,
  Phys.\ Rev.\  {\bf 184}, 1760 (1969);
  C.~J.~Isham, A.~Salam and J.~A.~Strathdee,
  Phys.\ Lett.\ B {\bf 31}, 300 (1970);
  C.~J.~Isham, A.~Salam and J.~A.~Strathdee,
  Annals Phys.\  {\bf 62}, 98 (1971);  B.~Zumino,
  in Brandeis Univ. 1970, Lectures On Elementary Particles And Quantum Field Theory, Vol. 2, 437-500;
  D.~V.~Volkov,
  Fiz.\ Elem.\ Chast.\ Atom.\ Yadra {\bf 4}, 3 (1973);
  W.~A.~Bardeen, M.~Moshe and M.~Bander,
  Phys.\ Rev.\ Lett.\  {\bf 52} (1984) 1188.
\bibitem{Tomboulis:1988gw}  A.~Cappelli and A.~Coste,
  Nucl.\ Phys.\ B {\bf 314}, 707 (1989);
  E.~T.~Tomboulis,
  Nucl.\ Phys.\ B {\bf 329}, 410 (1990); A.~Schwimmer and S.~Theisen,
  Nucl.\ Phys.\ B {\bf 847}, 590 (2011).
\bibitem{Nicolis:2008in}
  A.~Nicolis, R.~Rattazzi and E.~Trincherini,
  Phys.\ Rev.\ D {\bf 79} (2009) 064036
  \hhref{0811.2197}.

\bibitem{fubini}
  S.~Fubini,
  Nuovo Cim.\  A {\bf 34}, 521 (1976).

\bibitem{Nicolis:2009qm} 
  A.~Nicolis, R.~Rattazzi and E.~Trincherini,
  JHEP {\bf 1005}, 095 (2010)
  [Erratum-ibid.\  {\bf 1111}, 128 (2011)]
  \hhref{0912.4258}.

\bibitem{Sundrum:2003yt}
  R.~Sundrum,
  \hhref{hep-th/0312212}.

   
\bibitem{Evans:2010bp}
  J.~A.~Evans, J.~Galloway, M.~A.~Luty and R.~A.~Tacchi,
 \hhref{1001.1361}.

\bibitem{Coleman:1973jx}
  S.~R.~Coleman and E.~J.~Weinberg,
  Phys.\ Rev.\  D {\bf 7}, 1888 (1973).


\bibitem{Maldacena:1997re}
  J.~M.~Maldacena,
  Adv.\ Theor.\ Math.\ Phys.\  {\bf 2} (1998) 231
  \hhref{hep-th/9711200}.
%
\bibitem{wittensb} Witten, Talk at ITP conference {\it 
New Dimensions in Field Theory and String Theory}, Santa Barbara
[\href{http://www.itp.ucsb.edu/online/susy\_c99/discussion}{http://www.itp.ucsb.edu/online/susy\_c99/discussion}]
%
\bibitem{Gubser:1999vj}
  S.~S.~Gubser,
  Phys.\ Rev.\ D {\bf 63} (2001) 084017
  \hhref{hep-th/9912001}.
%
\bibitem{Verlinde:1999fy}
  H.~L.~Verlinde,
  Nucl.\ Phys.\ B {\bf 580} (2000) 264
  \hhref{hep-th/9906182}.




\bibitem{Randall:1999ee}
  L.~Randall and R.~Sundrum,
  Phys.\ Rev.\ Lett.\  {\bf 83} (1999) 3370
  \hhref{hep-ph/9905221}.




\bibitem{Goldberger:1999uk}
  W.~D.~Goldberger and M.~B.~Wise,
  Phys.\ Rev.\ Lett.\  {\bf 83} (1999) 4922
  \hhref{hep-ph/9907447}.

\bibitem{Goldberger:1999un}
  W.~D.~Goldberger and M.~B.~Wise,
  Phys.\ Lett.\ B {\bf 475} (2000) 275
  \hhref{hep-ph/9911457}.


\bibitem{DeWolfe:1999cp}
  O.~DeWolfe, D.~Z.~Freedman, S.~S.~Gubser and A.~Karch,
  Phys.\ Rev.\ D {\bf 62} (2000) 046008
  \hhref{hep-th/9909134}.


\bibitem{Csaki:2000zn}
  C.~Csaki, M.~L.~Graesser and G.~D.~Kribs,
  Phys.\ Rev.\ D {\bf 63} (2001) 065002
  \hhref{hep-th/0008151}.

\bibitem{Rattazzi:2000hs}
  R.~Rattazzi and A.~Zaffaroni,
  JHEP {\bf 0104} (2001) 021
  \hhref{hep-th/0012248}.

\bibitem{Lewandowski:2001qp}
  A.~Lewandowski and R.~Sundrum,
  Phys.\ Rev.\ D {\bf 65} (2002) 044003
  \hhref{hep-th/0108025}.

\bibitem{Kofman:2004tk}
  L.~Kofman, J.~Martin and M.~Peloso,
  Phys.\ Rev.\ D {\bf 70} (2004) 085015
  \hhref{hep-ph/0401189}.

\bibitem{Goldberger:2007zk}
  W.~D.~Goldberger, B.~Grinstein and W.~Skiba,
  Phys.\ Rev.\ Lett.\  {\bf 100} (2008) 111802
  \hhref{arXiv:0708.1463}.

\bibitem{Konstandin:2010cd}
  T.~Konstandin, G.~Nardini and M.~Quiros,
  Phys.\ Rev.\ D {\bf 82} (2010) 083513
  \hhref{1007.1468}.

\bibitem{Grinstein:2011dq}
  B.~Grinstein and P.~Uttayarat,
  JHEP {\bf 1107} (2011) 038
  \hhref{1105.2370}.

\bibitem{Eshel:2011wz}
  Y.~Eshel, S.~J.~Lee, G.~Perez and Y.~Soreq,
  JHEP {\bf 1110} (2011) 015
 \hhref{1106.6218}.





\bibitem{Chacko:2012sy}
  Z.~Chacko and R.~K.~Mishra,
  \hhref{1209.3022}.

\bibitem{Bellazzini:2012vz}
  B.~Bellazzini, C.~Csaki, J.~Hubisz, J.~Serra and J.~Terning,
  \hhref{1209.3299}.

\bibitem{Bellazzini:2013fga}
  B.~Bellazzini, C.~Csaki, J.~Hubisz, J.~Serra and J.~Terning,
 \hhref{1305.3919}.




\bibitem{ArkaniHamed:2000ds}
  N.~Arkani-Hamed, M.~Porrati and L.~Randall,
  JHEP {\bf 0108} (2001) 017
  \hhref{hep-th/0012148}.

\bibitem{Girardello:1998pd}
  L.~Girardello, M.~Petrini, M.~Porrati and A.~Zaffaroni,
  JHEP {\bf 9812} (1998) 022





\bibitem{Arnowitt}
  R.~L.~Arnowitt, S.~Deser and C.~W.~Misner,
  in ``Gravitation, an introduction to current research,''
  L. Witten ed.,
  Wiley, New York, 1962.

\bibitem{Luty:2003vm}
  M.~A.~Luty, M.~Porrati and R.~Rattazzi,
  JHEP {\bf 0309} (2003) 029
  \hhref{hep-th/0303116}.

\bibitem{Gibbons:1976ue}
  G.~W.~Gibbons and S.~W.~Hawking,
  Phys.\ Rev.\ D {\bf 15} (1977) 2752.

\bibitem{Wald:1984rg}
  R.~M.~Wald,
  ``General Relativity,''
  Chicago, Usa: Univ. Pr. ( 1984) 491p



\bibitem{Charmousis:1999rg} 
  C.~Charmousis, R.~Gregory and V.~A.~Rubakov,
  Phys.\ Rev.\ D {\bf 62}, 067505 (2000)
  \hhref{hep-th/9912160}.

\bibitem{Gubser:1998bc}
  S.~S.~Gubser, I.~R.~Klebanov and A.~M.~Polyakov,
  Phys.\ Lett.\ B {\bf 428} (1998) 105
 \hhref{hep-th/9802109}.
%
\bibitem{Witten:1998qj}
  E.~Witten,
  Adv.\ Theor.\ Math.\ Phys.\  {\bf 2} (1998) 253
  \hhref{hep-th/9802150}.




\bibitem{Kaloper:1999sm}
  N.~Kaloper,
  Phys.\ Rev.\ D {\bf 60} (1999) 123506
  \hhref{hep-th/9905210}.




  
\bibitem{Kiritsis:2006ua}
  E.~Kiritsis and F.~Nitti,
  Nucl.\ Phys.\ B {\bf 772} (2007) 67
  \hhref{hep-th/0611344}.

\bibitem{Lesgourgues:2003mi} 
  J.~Lesgourgues, L.~Sorbo and ,
  Phys.\ Rev.\ D {\bf 69}, 084010 (2004)
  \hhref{hep-th/0310007}.


\bibitem{'tHooft:1979bh} 
  G.~'t Hooft,
  NATO Adv.\ Study Inst.\ Ser.\ B Phys.\  {\bf 59}, 135 (1980).
  



\bibitem{SundrumInPreparation}
 R. Sundrum, in preparation. See also talk given at Stanford University, May 19 2012, \\
\href{http://www.stanford.edu/dept/physics/events/2012/SavasFest/slides/Raman\%20Sundrum.pdf}{http://www.stanford.edu/dept/physics/events/2012/SavasFest/slides/Raman\%20Sundrum.pdf}

\bibitem{Weinberg:1988cp} 
  S.~Weinberg,
  Rev.\ Mod.\ Phys.\  {\bf 61}, 1 (1989).

\bibitem{Brown:1988kg} 
  J.~D.~Brown and C.~Teitelboim,
  Nucl.\ Phys.\ B {\bf 297}, 787 (1988).

\bibitem{Feng:2000if} 
  J.~L.~Feng, J.~March-Russell, S.~Sethi and F.~Wilczek,
  Nucl.\ Phys.\ B {\bf 602}, 307 (2001)
  \hhref{hep-th/0005276}.

\bibitem{Koivisto:2009fb} 
  T.~S.~Koivisto and N.~J.~Nunes,
  Phys.\ Rev.\ D {\bf 80}, 103509 (2009)
  \hhref{arXiv:0908.0920}.













}
\end{thebibliography}
\end{document}